\documentclass[pra,twocolumn,longbibliography]{revtex4-2}
\usepackage[T1]{fontenc}
\usepackage{amsthm}
\usepackage{amsmath}
\usepackage{siunitx}
\usepackage[version=4]{mhchem}
\usepackage{datetime}

\usepackage{color}
\usepackage{newtxtext,newtxmath}
\usepackage{microtype}
\usepackage{braket}
\usepackage{xr}
\usepackage{natbib}
\usepackage{hyperref}
\usepackage{mathtools}
\usepackage{dsfont}
\usepackage[normalem]{ulem}
\usepackage{graphicx}

\hypersetup{%
    bookmarksopen=false,
    bookmarksnumbered=true,
    pdfnewwindow=true,
    unicode=false,
    colorlinks=true,%
    citecolor=blue,
    linkcolor=black,
    urlcolor=blue,
    filecolor=blue
    }
\graphicspath{{./Figures/}} 

\newcommand{\dket}[1]{|#1\rangle\!\rangle} % double ket
\renewcommand{\i}{\mathrm{i}}

\renewcommand{\vec}[1]{\mathbf{#1}}

\theoremstyle{definition}

\makeatletter
\newcommand*\bigcdot{\mathpalette\bigcdot@{.5}}
\newcommand*\bigcdot@[2]{\mathbin{\vcenter{\hbox{\scalebox{#2}{$\m@th#1\bullet$}}}}}
\makeatother

\begin{document}
%%%%%%%%%%%%%%%%%%%%%%%%%%%%%%%%%%%%%%%%%%%%%%%%%%%%%%%%%%%%%%%%%%%%%%%%%%
%%%affiliations
\newcommand{\liege}{Institut de Physique Nucléaire, Atomique et de Spectroscopie, CESAM, Universit\'e de Li\`ege, 4000 Liège, Belgium}

\title{The role of non-Markovian dissipation in quantum phase transitions: \\ 
tricriticality, spin squeezing, and directional symmetry breaking}

\author{Baptiste Debecker}
\email{Baptiste.Debecker@uliege.be}
\affiliation{\liege}
\author{Lukas Pausch}
\altaffiliation[Present address: ]{German Aerospace Center, Institute of Quantum Technologies, Ulm, Germany}
\affiliation{\liege}
\author{Jonathan Louvet}
\affiliation{\liege}
\author{Thierry Bastin}
\affiliation{\liege}
\author{John Martin}
\affiliation{\liege}
\author{François Damanet}
\affiliation{\liege}
\begin{abstract}
	Understanding how to control phase transitions in quantum systems is at the forefront of research for the development of new quantum materials and technologies. Here, we study how the coupling of a quantum system to a non-Markovian environment, i.e., an environment with a frequency-dependent spectral density inducing memory effects, can be used to generate and reshape phase transitions and squeezing in matter phases. Focusing on a Lipkin-Meshkov-Glick model, we demonstrate that non-Markovian dissipation can be leveraged to engineer tricriticality via the fusion of $2^{\mathrm{nd}}$-order and $1^{\mathrm{st}}$-order critical points. 
    We identify phases that arise from different ways of breaking the single weak symmetry of our model, which led us to introduce the concept of \textit{directional spontaneous symmetry breaking} (DSSB) as a general framework to understand this phenomenon.   
    We show that signatures of DSSB can be seen in the emergence of spin squeezing along different directions, and that the latter is controllable via non-Markovian effects, opening up possibilities for applications in quantum metrology. Finally, we propose an experimental implementation of our non-Markovian model in cavity QED. Our work features non-Markovianity as a resource for controlling phase transitions in general systems, and highlights shortcomings of the Markovian limit in this context.
\end{abstract}

\date{\today}
\maketitle

\section{Introduction}

Dissipative mechanisms have emerged as powerful tools for shaping the quantum states and phases of many-body systems. Rather than viewing the inevitable coupling of a quantum system to its environment purely as a source of decoherence, one can exploit dissipation to drive a system towards non-trivial steady states~\cite{Roberts2020D, Zapletal2022S}, engineer phases of matter without an equilibrium counterpart~\cite{Toner1998Flocks, Jin2016Cluster, Lee2013Unconventional}, induce quantum many-body correlations~\cite{Diehl2008Q, Verstraete2009Q} or design improved quantum metrology protocols~\cite{Koppenhofer2022D, Yang2024C}.

Most of the theoretical efforts in understanding the dissipative phases of matter have focused on Markovian dissipation, where the environment has no memory of its past interactions with the system. Although this approximation holds well in platforms when the system-environment coupling is weak and the bath correlation time is short~\cite{deVega2017Dynamics}, it can fail in many realistic settings, such as opto-mechanical~\cite{Groblacher2015} or light harvesting systems~\cite{Potocnik2018}, micropillar cavities~\cite{Madsen2011}, or photonic lattices~\cite{Vicencio2025}. In such cases, non-Markovian effects, i.e., the backflow of information from the environment to the system, become crucial. Moreover, non-Markovianity naturally emerges when one builds effective reduced descriptions of larger Markovian systems to decrease the computational complexity associated with large system sizes~\cite{Palacino2020, Damanet2019, Link2022}. Interestingly, recent work has shown that non-Markovian effects can lead to steady states inaccessible with Markovian generators~\cite{Ask2022}, enhance entanglement in steady states~\cite{Huelga2012Non}, considerably influence transport properties~\cite{Maier2019}, stabilize the skin effect from additional noise~\cite{Kuo2024N}, and reveal phases of matter hidden by the Markovian approximation~\cite{Debecker2024S, Debecker2024C}. The latter raises the question of whether non-Markovianity can be harnessed as a resource to discover and stabilize new phases of matter with tailored properties. 

In this work, we address that question by studying a non-Markovian generalization of the anisotropic Lipkin-Meshkov-Glick (LMG) model with a transverse magnetic field. This model was investigated in~\cite{Debecker2024C}, where it was shown that the non-Markovian LMG model can exhibit two consecutive second-order dissipative phase transitions (DPTs) separating three different phases, in stark contrast to the unique trivial infinite temperature steady state found in the Markovian limit. Here, we demonstrate that non-Markovianity can be used as a tool to merge these two second-order DPTs into a first-order DPT, thus establishing the existence of a tricritical point. Furthermore, we interpret tricriticality through the lens of a new symmetry breaking concept that we call \textit{Directional Spontaneous Symmetry Breaking} (\textit{DSSB}) and show how it provides a unifying picture of the mechanism leading to different flavors of symmetry breaking, in our model, but also in other known tricritical models~\cite{Soriente2018D}. In addition, we calculate exactly a spin squeezing parameter in all regimes. We show how it highlights the concept of DSSB, via the emergence of squeezing along orthogonal directions depending on the parameter region in the phase diagram, and demonstrate that it cannot serve as a simple phase identifier: \emph{within} a given phase, the memory of the environment can be adjusted to generate or enhance spin squeezing, which indicates that non-Markovianity can therefore also serve as a resource for the generation of steady-state entanglement. Finally, we propose a scheme relevant for current cavity QED platforms to experimentally realize the non-Markovian LMG model that we investigate theoretically, thereby offering a direct pathway to explore quantum phase transitions in driven-dissipative systems. 

This paper is organized as follows: In Sec.~II, we present our generalized LMG model and its symmetry. In Sec.~III, we discuss the emergence of dissipative phase transitions in the model and their connection with the spontaneous breaking of its symmetry. To do this, we first perform a mean-field analysis valid in the thermodynamic limit, which allows us to construct a phase diagram; then we present the concept of directional spontaneous symmetry breaking, providing a refined understanding of the breaking of the symmetry, and finally, we show the finite-size behavior of the system. In Sec.~IV, we analyze the spin squeezing parameter and show how it deviates in some regimes from the phase identification. In Sec.~V, we propose an experimental cavity QED setup that implements our model. In Sec.~VI, we conclude and provide some perspectives on our work.

\section{Non-Markovian LMG Model} \label{sec::Model}
We consider a system composed of $N$ spin-1/2 particles that interact through infinite-range interactions and are embedded in a magnetic field. The system's Hamiltonian $H_S$ is given by (we set $\hbar=1$)
\begin{equation}
    H_{\mathrm{S}} = \frac{V}{N}(S_x^2 - S_y^2) + h S_z, 
    \label{H_S}
\end{equation}
where $V$ controls the spin-spin interaction, $h$ characterizes the intensity of the magnetic field applied in the $z$ direction, and $S_i = \tfrac{1}{2} \sum_{j=1}^{N} \sigma_i^{(j)}$ ($i=x, y, z$) are collective spin operators expressed through single-spin Pauli operators $\sigma_i^{(j)}$. This Hamiltonian is a specific realization of the paradigmatic Lipkin-Meshkov-Glick (LMG) model, introduced in~\cite{Lipkin1965V}, which has been extensively studied (see, e.g. Refs~\cite{Lipkin1965V, Vidal2004E, Wilms2012J, Pedro2007T, Orus2008E, Ptaszy2024D}). 

In this work, we extend the standard LMG model to an open system framework, accounting for the interaction between the spins and an environment modeled as an infinite collection of bosonic modes. The total Hamiltonian $H_\mathrm{tot}$ for the system-environment couple is
\begin{equation}
\begin{split}
    &H_\mathrm{tot} = H_{\mathrm{S}} + H_{\mathrm{E}} + H_\mathrm{int}, \\
    &H_{\mathrm{E}} = \sum_k \omega_k a_k^\dagger a_k, \quad H_\mathrm{int} = \sum_k (g_k S_x a_k+ g_k^* S_x a_k^\dagger),
    \label{H_SE}
\end{split}
\end{equation}
 where $H_{\mathrm{E}}$ ($H_\mathrm{int}$) is the environment (interaction) Hamiltonian, $a_k$ ($a_k^\dagger$) is the annihilation (creation) operator of the bosonic mode $k$ of frequency $\omega_k$ and $|g_k|$ is the interaction strength between the system and mode $k$. We focus exclusively on the spin degrees of freedom at zero temperature whose dynamics is entirely determined by the vacuum bath correlation function 
\begin{equation}
 \begin{split}
    \alpha(t-s) &\equiv \sum_k |g_k|^2 \braket{a_k(t) a_k^\dagger(s)} = \sum_k |g_k|^2 e^{-i\omega_k (t-s)}, \\
    \end{split}
    \label{defCF}
\end{equation}
with $a_k(t) = a_k e^{-i\omega_k t}$ the annihilation operator in the interaction picture with respect to $H_{\mathrm{E}}$. 

A commonly studied case is the uniform weak-coupling scenario, where the spectral density $J(\omega) \equiv \sum_k |g_k|^2 \delta(\omega-\omega_k)$ is assumed to be a constant
\begin{equation}
    J(\omega) \approx \frac{\gamma}{N \pi},
\end{equation}
in the continuum limit of modes. This requires performing approximations on the real environment of the system and their coupling, and it greatly simplifies the correlation function in the continuum limit, which reads
\begin{equation}
 \begin{split}
 \alpha(t-s)
 &= \int_0^{+\infty} d\omega~J(\omega) e^{-i\omega (t-s)}\\
 &\approx \frac{\gamma}{N\pi} \int_{0}^{+\infty} d\omega~   e^{-i \omega (t-s)} \\
& = \frac{\gamma}{N\pi} \left[ \pi \delta(t-s) + i\, \mathrm{P.V.}\left(\frac{1}{t-s}\right) \right] \\
&\approx \frac{\gamma}{N} \delta(t-s).
 \end{split}
\end{equation}
Here, $\mathrm{P.V.}$ denotes the Cauchy principal value, which we have discarded following standard practice, as it ultimately leads to a coherent part that amounts to renormalize the energy of the system. The delta-distribution correlation function implies that the environment lacks memory of its interaction with the system. Physically, this corresponds to the so-called Markovian limit, where the interaction of the system with a vast and unstructured environment does not result in any feedback from the environment to the system, ensuring a memoryless evolution~\cite{Bre06}. In this regime, one can readily derive a Lindblad master equation for the system density operator $\rho_S$ only, which takes the form~\cite{Lindblad1976b, Gorini1976}
\begin{equation}
\label{MarkovLimit}
\dot{\rho}_\mathrm{S} = -i[H_{\mathrm{S}}, \rho_\mathrm{S}] + {\frac{\gamma}{2N}}\left(2 S_x\rho_\mathrm{S} S_x^\dagger - \{ S_x^\dagger S_x, \rho_\mathrm{S}\} \right).
\end{equation}
Similar open LMG models have previously been investigated in the Markovian regime~\cite{Pausch2024, FerreiraL2019, Ma2020P, Lee2014D, Morrison2008D}. In contrast to the ground state of the Hamiltonian~\eqref{H_S}, which exhibits quantum criticality and spin-squeezing~\footnote{Given the quadratic form of the Hamiltonian \eqref{H_S}, the ground state in the thermodynamic limit $N \to \infty$ is Gaussian and therefore a coherent or squeezed state, since it is pure. It is coherent only when $V/h = 0$, as a spin-coherent state can only be an eigenstate of \eqref{H_S} when $H_{\mathrm{S}} = hJ_z$. For $V/h \neq 0$, the state is therefore squeezed.}, the steady state of the master equation~\eqref{MarkovLimit} is not critical, as $\rho_\mathrm{ss} \propto \mathds{1}$ is the unique \cite{Nigro2019Uniqueness} steady state for all $N$. Otherwise stated, the quantum criticality hosted by the LMG Hamiltonian is suppressed when the dissipation is assumed to be Markovian. 

Here, we go beyond the idealized Markovian scenario and consider a structured environment with a correlation function of the form 
\begin{equation}
   \alpha(t-s) =  \frac{\gamma \kappa}{2N}e^{-i\omega |t-s| - \kappa |t-s|}, 
   \label{CF}
\end{equation}
This choice is not only experimentally relevant [especially for cavity or circuit QED setups~\cite{Morrison2008C, Mivehvar2021Cavity, Blais2021Circuit} (see also Sec.\ref{section:exp})], but turns out to be especially convenient. Firstly, in the limit $\kappa \rightarrow + \infty $, Eq.~\eqref{CF} reduces to $\alpha(t-s) = (\gamma/N) \delta(t-s)$, recovering the Markovian limit and thus the master equation~(\ref{MarkovLimit}) for the system density operator. Secondly, the limit $\kappa \rightarrow 0$ simply yields $\alpha(t-s) = 0$: the environment decouples from the system, and we recover the closed version of the LMG model~(\ref{H_S}). Finally, for finite $\kappa \ne 0$, the dynamics of the system is non-Markovian. However, for these dynamics with a correlation function of the form~\eqref{CF}, the pseudomode picture~\cite{Imamoglu1994Stochastic, Dalton2001Theory, Garraway1997Nonperturbative, Pleasance2020Generalized, Mazzola2009Pseudomodes, Yang2012Nonadiabatic, Breuer2004Genuine, Barchielli2010Stochastic} provides a Markovian embedding described by a Markovian master equation of the form
\begin{equation} 
\dot{\rho}_\mathrm{S+a} = -i[H, \rho_\mathrm{S+a}] + \kappa\left(2 a \rho_\mathrm{S+a} a^\dagger - \{ a^\dagger a, \rho_\mathrm{S+a}\} \right),
   \label{master_LMG_tot}
\end{equation}
with
\begin{equation}
    H = H_{\mathrm{S}} + \sqrt{\frac{\gamma \kappa}{2N}}S_x(a+a^\dagger) + \omega a^\dagger a,
    \label{H_LMG_tot}
\end{equation}  
where $a$ is the pseudomode annihilation operator and $\rho_{S+a}$ the density operator for the system and pseudomode. The spin state $\rho_S$ can be obtained exactly by tracing out the pseudomode $a$. Therefore, to extract finite-size results, one could in principle solve Eq.~(\ref{master_LMG_tot}) for $\rho_\mathrm{S+a}$ and trace out the $a$ mode in order to study the non-Markovian system made up of spins only. However, this approach is usually more costly numerically~\cite{Debecker2024S} and we choose instead to employ the framework that has been recently developed in the context of non-Markovian dissipative phase transitions~\cite{Debecker2024S, Debecker2024C} and which is based on Hierarchical Equations of Motion (HEOM)~\cite{Tanimura2020, Tanimura89} described below in Subsec.~\ref{sec:HEOM}. Note that the master equation~\eqref{master_LMG_tot} will prove useful to derive exact results in the thermodynamic limit $N \to \infty$.

The master equation~\eqref{master_LMG_tot} can be rewritten as
\begin{equation}
\begin{split}
    \dot{\rho}_\mathrm{S+a} &= \mathcal{L}_M[\rho_\mathrm{S+a}], \quad \mathcal{L}_M = \mathcal{H} + \mathcal{D} \\
    \mathcal{H}[\cdot] &= -i[H, \cdot], \quad \mathcal{D}[\cdot] = \kappa(2a\cdot a^\dagger - \{a^\dagger a, \cdot\}),
    \label{def_LM}
 \end{split}
\end{equation}
where the superoperator $\mathcal{L}_M$ is the so-called Liouvillian, Lindbladian, or generator of the dynamics, and $\mathcal{H}$ ($\mathcal{D}$) is the superoperator that dictates the coherent (dissipative) dynamics. The Liouvillian is invariant under the transformation
\begin{equation}\label{rep}
    a \rightarrow -a, \quad S_x \rightarrow -S_x, \quad S_y \rightarrow -S_y,
\end{equation}
described by the \textit{unitary} superoperator 
\begin{equation}
   \mathcal{U}[\cdot] = U \cdot U^\dagger, \quad U = e^{i\pi(S_z + a^\dagger a)}
   \label{def_Z2}
\end{equation}
satisfying
\begin{equation}
    [\mathcal{L}_M, \mathcal{U}] = 0.
    \label{weak_sym_def}
\end{equation}
Unitary superoperators verifying~\eqref{weak_sym_def} are called \textit{weak symmetries} throughout the literature~\cite{Buca2012}. Since $\mathcal{U} ^2 = \mathds{1}$, $\mathcal{U}$ is a weak $\mathbb{Z}_2$-symmetry. Note that the existence of such a symmetry does not imply a conservation law~\cite{Albert2014S}. However, it does imply that, if the steady state is unique (which is the case in our model for any \textit{finite} $N$), then, in the steady state, $\braket{a} = \braket{S_x} = \braket{S_y} = 0$ and the spin expectation value is in the $z$ direction. A finer discussion of the model's symmetries is presented in Subsec.~\ref{Sec:DSSB}.

\section{Dissipative phase transitions}
\label{sec:DPT}
We now analyze the existence of dissipative phase transitions in our model. After providing a definition of dissipative phase transitions, we present a mean-field analysis, allowing the construction of a complete phase diagram (shown in Fig.~\ref{fig:PhaseDiagrams}). Then, as the latter contains two phases consisting of two different ways of breaking the symmetry of the model, we introduce the concept of DSSB to achieve a deeper understanding of the underlying mechanism of tricriticality. Finally, we present a finite-size scaling analysis of the emergence of criticality and study the finite-size consequences of the DSSB.
\\

\subsection{Definition} 

We define a dissipative phase transition (DPT) in our model by the existence of a system observable $O$, independent of the parameter $V$, whose steady-state expectation value displays a nonanalytic behavior in the thermodynamic limit $N \rightarrow +\infty$. Formally, this definition can be written as~\cite{Minganti2018Spectral}
\begin{equation}\label{defDPT}
     \lim_{V\to V_i} \left|\frac{\partial^p}{\partial V^p} \lim_{N\to +\infty} \langle O \rangle_\mathrm{ss}\right| = +\infty,
\end{equation}
where $\langle O \rangle_\mathrm{ss}$ denotes the steady-state expectation value of $O$ and $p$ is the order of the transition. In what follows, we will mainly deal with expectation values of collective spin operators and DPTs of orders $p = 1$ and $p = 2$.

\subsection{Mean-field analysis} \label{section::MF}
In this section, we analyze the steady-state behavior of our model in the thermodynamic limit $N \to \infty$, where a mean-field approach is exact~\cite{Hepp1973O}. The mean-field equations of motion for $\langle a\rangle$, $\langle S_x\rangle$, $\langle S_y \rangle $, and $\langle S_z \rangle $ can be obtained from Eqs.~(\ref{master_LMG_tot}) and (\ref{H_LMG_tot}) by assuming $\langle A B \rangle = \langle A\rangle \langle B \rangle + \tfrac12 \langle [A, B] \rangle$ for any operator $A$ and $B$. They read 
\begin{align}\label{EOMFull_12}
     \langle \dot a \rangle &= -(\kappa + i \omega) \langle a \rangle - i \sqrt{\frac{\gamma \kappa}{2 N}} \langle S_x \rangle, \\ \label{EOMFull22}
    \langle \dot S_x \rangle &= - 2\frac{V}{N} \langle S_y \rangle \langle  S_z \rangle - h \braket{S_y},\\\label{EOMFull32}
    \langle \dot S_y \rangle &= - 2\frac{V}{N} \langle S_x \rangle \langle  S_z \rangle + h \braket{S_x} - \sqrt{\frac{\gamma \kappa}{2N}} \langle S_z\rangle  (\langle a \rangle+ \langle a^\dagger \rangle),\\\label{EOMFull42}
    \langle \dot S_z \rangle &= 4\frac{V}{N} \langle S_x \rangle \langle  S_y \rangle +  \sqrt{\frac{\gamma \kappa}{2N}} \braket{S_y} (\braket{a} + \braket{a^\dagger}).
\end{align}
By setting the left-hand side of the equations above to zero, we obtain (except for $V = \gamma \omega \kappa/[4(\kappa^2 + \omega^2)]$ as explained below) six fixed points from which four phases labeled (I), (IIa), (IIb) and (III) can be inferred:
\begin{widetext}
\begin{align}
&\text{(I): } \left(\langle a \rangle_\mathrm{ss}, \langle S_x \rangle_\mathrm{ss}, \langle S_y\rangle_\mathrm{ss}, \langle S_z\rangle_\mathrm{ss}  \right) = \frac{N}{2} \left(\mp \sqrt{\frac{\gamma}{2 N \kappa}}\sqrt{1 - \left(\frac{ h}{V - \frac{q_2 \gamma}{2}}\right)^2} (q_2 + i q_1), \pm \sqrt{1 - \left(\frac{ h}{V - \frac{q_2 \gamma}{2}}\right)^2},0, \frac{h}{V-\frac{q_2 \gamma}{2}}\right) \label{MF-phaseI}\\
&\text{(IIa): } \left(\langle a \rangle_\mathrm{ss}, \langle S_x \rangle_\mathrm{ss}, \langle S_y\rangle_\mathrm{ss}, \langle S_z\rangle_\mathrm{ss}  \right) = \frac{N}{2}\left(0,0,0, -1\right) \label{MF-phaseII} \\
&\text{(IIb): } \left(\langle a \rangle_\mathrm{ss}, \langle S_x \rangle_\mathrm{ss}, \langle S_y\rangle_\mathrm{ss}, \langle S_z\rangle_\mathrm{ss}  \right) = \frac{N}{2}\left(0,0,0, 1\right)\label{MF-phaseIIb} \\
&\text{(III): } \left(\langle a \rangle_\mathrm{ss}, \langle S_x \rangle_\mathrm{ss}, \langle S_y\rangle_\mathrm{ss}, \langle S_z\rangle_\mathrm{ss}  \right) =  \frac{N}{2}\left(0,0,\pm \sqrt{1 - \left(\frac{h}{V}\right)^2}, -\frac{h}{V}\right).
\label{MF-phaseIII}
\end{align}
\end{widetext}
where
\begin{equation}
    q_1 = \frac{\kappa^2}{\kappa^2 + \omega^2},\qquad q_2 = \frac{\kappa \omega}{\kappa^2 + \omega^2}
\end{equation}
are adimensional parameters that capture non-Markovian effects due to the spectral structure of the bath, and where the notation $\langle O \rangle_\mathrm{ss} \equiv \langle O(t \to\infty)\rangle$ denotes the steady state expectation value of an operator $O$. In the following, for simplicity, we omit the subscript $_\mathrm{ss}$ because we only consider steady-state expectation values (unless otherwise stated).\\

If $h > 0$ [$h < 0]$, the phase (IIb) [(IIa)] is always unstable. Focusing on the case $h > 0$ without loss of generality, we thus only have the phases (I), (IIa), and (III) to consider. Phase (IIa), which we will refer to  as phase (II) from now on, has a single fixed point, while phases (I) and (III) both exhibit two fixed points. The pair of fixed points in those phases corresponds to broken symmetry states that relate to each other via the transformation~(\ref{rep}). Note that, from a direct examination of Eqs.~(\ref{MF-phaseI}) and (\ref{MF-phaseIII}), the fixed points of phase (I) and (III) are unphysical for $| V - q_2\gamma/2| < h $ and $|V| < h$, respectively.

\begin{figure*}
    \centering
\includegraphics[width=0.975\textwidth]{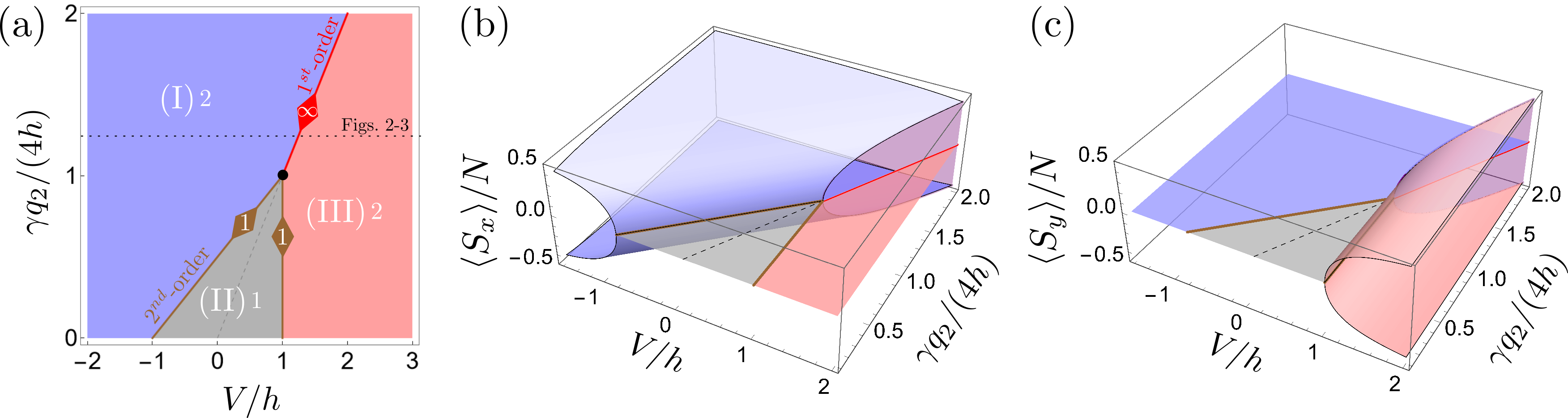}
\caption{\textbf{Mean-field analysis.} (a) Phase diagram obtained from the mean-field analysis of our model. The roman numerals (I) (II) and (III) label the different phases, while the arabic numerals (1, 2 and $\infty$) indicate the number of distinct stable fixed points within each phase and on the critical lines (see colored diamonds) that correspond either to 2$^{\mathrm{nd}}$-order (brown) or $1^{\mathrm{st}}$-order (red) dissipative phase transitions, respectively. The black dot indicates the position of the tricritical point, at $V = h = \gamma q_2/4$. The thin horizontal dashed black line indicates the parameters used in Figs.~\ref{Fig_sy2} and \ref{fig:DSSB}. (b) Possible steady state values of $\langle S_x \rangle$ in the different phases, highlighting the broken symmetry in phase (I) due to the existence of two possible opposite values of $\langle S_x \rangle$ but also the $1^{\mathrm{st}}$-order (2$^{\mathrm{nd}}$-order) nature of the transition (I)$\leftrightarrow$(III) [(I)$\leftrightarrow$(II)] from the jump (smooth behavior) of $\langle S_x \rangle$ across the red (brown) transition line. Note that at the critical line of the $1^{\mathrm{st}}$-order transition, an infinite set of possible values of $\langle S_x \rangle$ is available, as denoted by the light purple area. (c) Same for $\langle S_y \rangle$, highlighting the broken symmetry in phase (III) and the first- and second-order nature of the transitions (I)$\leftrightarrow$(III) and (II)$\leftrightarrow$(III), respectively.  Note that at the critical line of the $1^{\mathrm{st}}$-order transition, an infinite set of possible values of $\langle S_y \rangle$ is available, as denoted by the light purple area. } 
    \label{fig:PhaseDiagrams}
\end{figure*}

A standard linear stability analysis around the fixed points (see, e.g., Refs.~\cite{Damanet2019, Debecker2024C} for a detailed description of the procedure) allows us to determine which fixed points are stable in which parameter region and thus build the phase diagram shown in Fig.~\ref{fig:PhaseDiagrams}~(a), where we assumed $h, q_2,\gamma > 0$ for simplicity. 

For $\gamma < 4 h/q_2$, two critical points (brown lines in Fig.~\ref{fig:PhaseDiagrams}) appear upon varying $V$:
\begin{align}
    &V_1 = -h + \frac{q_2 \gamma}{2}, \\
    &V_2 = h.
\end{align} 
Phase (I) [blue] is stable for $V < V_1$. Phase (II) [gray] is stable for $V_1 < V < V_2$. Finally, Phase (III) [red] is stable for $V_2 < V$. In this regime, the transitions (I)$\leftrightarrow$(II) and (II)$\leftrightarrow$(III) are both of second order, as per the definition (\ref{defDPT}) with $p = 2$. Indeed, for the given transitions at $V_1$ and $V_2$, examples of observables $O$ satisfying Eq.~(\ref{defDPT}) are $S_x$ and $S_y$, respectively. They are shown in Fig.~\ref{fig:PhaseDiagrams}. For the transition (I)$\leftrightarrow$(II) [(II)$\leftrightarrow$(III)], $\langle S_x\rangle$ [$\langle S_y\rangle$] exhibits a (continuous) kink at $V_1$ [$V_2$], highlighting the second-order nature of the transition. 

For $\gamma \geq 4 h/q_2$, phase (II) no longer appears. In fact, the two critical points $V_1$ and $V_2$ of the two second-order DPTs coalesce at a tricritical point $\gamma = 4 h/q_2$ to emerge for $\gamma > 4 h/q_2$ as a single critical line (red line in Fig.~\ref{fig:PhaseDiagrams})
\begin{equation}\label{V3}
    V =  \frac{q_2 \gamma}{4}.
\end{equation}
Along this particular line (and whatever the value of $\gamma$), the mean-field equations exhibit an
infinite set of fixed points. Indeed, setting the left-hand side of equations~(\ref{EOMFull_12})-(\ref{EOMFull42}) to zero yields for $V =  q_2 \gamma/4$ a standard slaving of the bosonic mode to the collective spin 
\begin{equation}
    \langle a \rangle = -i \sqrt{\frac{\gamma \kappa}{2N}}\frac{1}{\kappa + i \omega}\langle S_x\rangle,
\end{equation}
as it appears when one performs the adiabatic elimination of the bosonic mode,
and the following remaining constraints (the equation for the last component simply reduces to $\langle \dot{S}_z\rangle = 0$)
 \begin{align}
    &0 = \left(\frac{\gamma q_2}{2N} \langle S_z\rangle + h \right) \langle S_y \rangle, \\
    &0 = \left(\frac{\gamma q_2}{2N} \langle S_z\rangle + h \right) \langle S_x \rangle,
\end{align}
in addition to the normalization condition $\langle S_x \rangle^2+\langle S_y \rangle^2+\langle S_z \rangle^2 = (N/2)^2$. 
If $\langle S_z \rangle \neq -2hN/\gamma q_2$, we recover the fixed point~(\ref{MF-phaseII}). If, however, $\langle S_z \rangle = -2hN/\gamma q_2$, then $\langle S_x \rangle$ and $\langle S_y \rangle$ are only constrained by the normalization condition. They can be parametrized as $\langle S_x \rangle = S \cos\theta$ and $\langle S_y \rangle = S \sin\theta$ with $S = (N/2)\sqrt{1-(4h/(\gamma q_2))^2}$ and $\theta\in [0,2\pi[$. This yields a whole family of fixed points of the form
\begin{equation}
\begin{aligned}
&\left(\langle a \rangle, \langle S_x \rangle, \langle S_y\rangle, \langle S_z\rangle  \right) = \\ 
&\frac{N}{2}\left(-\sqrt{\frac{\gamma}{2N \kappa}}(q_2 + i q_1)S \cos\theta, S\cos\theta,S \sin\theta, -\frac{4h}{\gamma q_2}\right).\label{MF-phaseIV}
\end{aligned}
\end{equation}
For $\gamma \leq 4 h/q_2$, this infinite set of fixed points is unstable (only the fixed point~(\ref{MF-phaseII}) is stable in this region of parameter, i.e., within phase (II), as expected). They are stable only for $\gamma > 4 h/q_2$, i.e., along the red critical line. Thus, there is a continuous $\mathbb{U}(1)$-symmetry of the spin steady state in the thermodynamic limit around the $z$-axis on the critical line, which is spontaneously broken for $\gamma > 4 h/q_2$ and thus leads to an infinite set of possible mean-field solutions for $\langle S_x \rangle$ and $\langle S_y \rangle$ [see light purple area in Fig.~\ref{fig:PhaseDiagrams}~(b) and (c)]. Furthermore, as one approaches the critical line, the values of $\braket{S_x}$ (or $\braket{S_y}$) from below do not match those from above. According to Eq.~(\ref{defDPT}), this mismatch in limits signals a first-order phase transition. In Subsec.~\ref{subsec::DPTandSpinSqueezing}, we will show that the emerging $\mathbb{U}(1)$ symmetry impacts the spin squeezing parameter in the thermodynamic limit.

\subsection{Directional spontaneous symmetry breaking}\label{Sec:DSSB}

Dissipative phase transitions and spontaneous symmetry breaking are intimately connected and have been extensively studied in the literature (see, e.g.,~\cite{Minganti2018Spectral, Minganti2021Continuous, Huber2020, Gu2024S, Debecker2024C}). In this section, we first review how conventional $\mathbb{Z}_2$-symmetry breaking arises, and then introduce the notion of Directional Spontaneous Symmetry Breaking (DSSB), which sheds new light on the phase diagram in Fig.~\ref{fig:PhaseDiagrams}.
\subsubsection{Spontaneous symmetry breaking}
Consider the Liouvillian $\mathcal{L}_M$ [Eq.~~\eqref{def_LM}] and its associated \newline $\mathbb{Z}_2$-symmetry superoperator $\mathcal{U}$ introduced in Eq.~\eqref{def_Z2}. Due to its simple $\mathbb{Z}_2$ nature, the superoperator $\mathcal{U}$ has only two eigenvalues: $u_{k} = \exp(ik\pi)~(k =0, 1)$, which defines two symmetry sectors, namely the symmetry sector $k=0$ (associated with $u_0 = 1$) and $k=1$ (associated with $u_1 = -1$). Since $[\mathcal{U}, \mathcal{L}_M] = 0$, the Liouvillian block-diagonalizes as
\begin{equation}
    \mathcal{L} = \mathcal{L}_{1} \oplus \mathcal{L}_{-1},
\end{equation}
where the block $\mathcal{L}_{u_{k}}$ is associated with the eigenvalue $u_{k}$. Because of this block-diagonal structure, the Liouvillian cannot mix different symmetry sectors. Furthermore, the steady state $\rho_\mathrm{ss}$ (assumed to be unique) always belongs to the symmetry sector $k=0$, i.e., $\mathcal{U}\rho_\mathrm{ss} = \rho_\mathrm{ss}$~\cite{Minganti2018Spectral}.

If, in each symmetry sector labeled by $k$, we denote the eigenvalues by $\lambda_j^{(k)}$ ($j = 0,1,\dotsc$) and sort them by increasing absolute value of their real part $|\mathrm{Re}[\lambda_0^{(k)}]| < |\mathrm{Re}[\lambda_1^{(k)}]| < \dotsc$, then a SSB is simply the emergence, in some region of the parameter space, of a zero eigenvalue in the sector $k=1$ in the thermodynamic limit, i.e., $\lambda_0^{(1)} \rightarrow 0$ as $N\rightarrow +\infty$. Consequently, the null space becomes degenerate and steady states that explicitly break the symmetry emerge. 

In the parameter regions where the symmetry is broken, one can show~\cite{Minganti2018Spectral} that the eigenoperator $\rho_0^{(1)}$ associated with the eigenvalue $\lambda_0^{(1)}$ is Hermitian. The density matrices 
\begin{equation}
    \rho_\pm =  \rho_0^{(0)} \pm \rho_0^{(1)},
    \label{rho_pm}
\end{equation}
with $\mathrm{Tr}[\rho_0^{(0)}] = 1$ and $\mathrm{Tr}[\rho_0^{(1)}] = 0$, are valid steady states that break the symmetry since $\mathcal{U}\rho_\pm = \rho_\mp$.

\subsubsection{Antiunitary flips}

As shown in Fig.~\ref{fig:PhaseDiagrams}~(b) and in Fig.~\ref{fig:PhaseDiagrams}~(c), and as already revealed by the fixed points~\eqref{MF-phaseI} and~\eqref{MF-phaseIII}, the system can break its underlying $\mathbb{Z}_2$ symmetry in two distinct ``directions'': 
\begin{enumerate}
    \item Breaking $S_x \rightarrow -S_x$ and $a \rightarrow -a$ simultaneously, while preserving $S_y \rightarrow -S_y$ [leading to phase (I)]. 
    \item Breaking $S_y \rightarrow -S_y$ while preserving $S_x \rightarrow -S_x$ and $a \rightarrow -a$ [leading to phase (III)].
\end{enumerate}
    
We formalize these two routes by introducing the two corresponding antiunitary operators $\mathcal{T}_I$ and $\mathcal{T}_\mathrm{III}$ defined by
\begin{equation}
  \mathcal{T}_\mathrm{I}[S_x] = -S_x,~\mathcal{T}_\mathrm{I}[a] = -a,~\mathcal{T}_\mathrm{I}[A] = A  \quad (A = S_y, S_z),
\end{equation}
 and 
\begin{equation}
  \mathcal{T}_\mathrm{III}[S_y] = -S_y,\quad\mathcal{T}_\mathrm{III}[A] = A \quad (A = S_x, S_z, a).
\end{equation}
 Denoting the complex conjugation in the $S_z$ basis by $\bigcdot^*$, the two antinunitary operators can be explicitly written as 
\begin{equation}
    \mathcal{T}_\mathrm{I}[\bigcdot] = U \bigcdot^* U^\dagger, \quad \mathcal{T}_\mathrm{III}[\bigcdot] = \bigcdot^*.
\end{equation}
 In the following, we restrict the action of the directional antiunitary flips to the (real) Hilbert space of Hermitian matrices (as they typically act on density matrices). One can readily verify that each operator is an involution ($\mathcal{T}_i^2=\mathds{1}$ for $i=\mathrm{I},\mathrm{III}$). They commute with each other, and anticommute (commute) with the coherent (dissipative) part of the Liouvillian, i.e., 
\begin{equation}
    [\mathcal{T}_\mathrm{I},\mathcal{T}_\mathrm{III}] = 0,\quad \{\mathcal{H}, \mathcal{T}_i \} = 0, \quad 
    [\mathcal{D}, \mathcal{T}_i] = 0\quad (i= \mathrm{I}, \mathrm{III}).
\end{equation}
We stress that neither $\mathcal{T}_\mathrm{I}$ nor $\mathcal{T}_\mathrm{III}$ commutes with the Liouvillian $\mathcal{L}_M$, in contrast to similar models that exhibit anitunitary symmetries~\cite{Pausch2024}. Nevertheless, their product 
\begin{equation}\label{Udecomp}
    \mathcal{T}_\mathrm{I}\, \mathcal{T}_\mathrm{III} = \mathcal{U}
\end{equation}
turns out to be the unitary $\mathbb{Z}_2$-symmetry $\mathcal{U}$ that does commute with $\mathcal{L}_M$. Therefore, we can interpret $\mathcal{U}$ as being generated by two \textit{directional flips} $\mathcal{T}_\mathrm{I}$ and $\mathcal{T}_\mathrm{III}$, each flipping a different subset of the model's degrees of freedom.

 Crucially, since $\mathcal{T}_\mathrm{I}$ and $\mathcal{T}_\mathrm{III}$ commute, we can build common eigenoperators $A_{\alpha_\mathrm{I} \alpha_\mathrm{III}}$ such that 
\begin{equation}
    \mathcal{T}_i[A_{\alpha_\mathrm{I} \alpha_\mathrm{III}}]  = \alpha_i A_{\alpha_\mathrm{I} \alpha_\mathrm{III}}, %\quad  \mathcal{T}_\mathrm{III}[A_{\alpha_\mathrm{I}, \alpha_\mathrm{III}}]= \alpha_\mathrm{III} A_{\alpha_\mathrm{I}, \alpha_\mathrm{III}},
    \label{A_alpha}
\end{equation}
with $\alpha_i = \pm 1~(i= \mathrm{I}, \mathrm{IIII})$. Consequently, the eigenvalues of the unitary symmetry $\mathcal{U}$ can be written as $\alpha_\mathrm{I} \alpha_\mathrm{III}$. 

\subsubsection{From SSB to DSSB} \label{subsubsec:fromSSBtoDSSB}
We now formally introduce the concept of Directional Spontaneous Symmetry Breaking (DSSB). In the following, we always assume the thermodynamic limit and we work in regions where $\mathcal{U}$ is broken (phases I and III). 

The broken symmetry steady states $\rho_\pm$ introduced in Eq.~\eqref{rho_pm} can be written as a real linear combination 
\begin{equation}
    \rho_\pm = \underbrace{a\,\rho_{++} + b\,\rho_{--}}_{\rho_0^{(0)}}\pm  \underbrace{( c\,\rho_{+-} + d\,\rho_{-+})}_{\rho_0^{(1)}},
\end{equation}
where the Hermitian joint eigenoperators $\rho_{\alpha_\mathrm{I} \alpha_\mathrm{III}}$ satisfy $\mathcal{T}_I [\rho_{\alpha_\mathrm{I} \alpha_\mathrm{III}}] = \alpha_\mathrm{I} \rho_{\alpha_\mathrm{I} \alpha_\mathrm{III}}$ and $\mathcal{T}_\mathrm{III} [\rho_{\alpha_\mathrm{I} \alpha_\mathrm{III}}] = \alpha_\mathrm{III} \rho_{\alpha_\mathrm{I}\alpha_\mathrm{III}}$. Furthermore, $\rho_0^{(0)}$ and $\rho_0^{(1)}$ must be orthogonal, as they live in different eigenspaces of the Hermitian operator $\mathcal{U}$. Thus, $\rho_\pm$ must also be orthogonal to each other, which gives the orthogonality condition
\begin{equation}
    \big\langle\rho_0^{(0)} \big|\rho_0^{(1)}\big\rangle_\mathrm{HS} = 0 \;\;\Leftrightarrow\;\; a^2 + b^2 =c^2 + d^2,
\end{equation}
with $\braket{A|B}_\mathrm{HS} \equiv \mathrm{Tr}[A^\dagger B]$ the standard Hilbert-Schmidt inner product.

Now, we consider the different possible routes for spontaneous symmetry breaking. The breaking of $\mathcal{T}_\mathrm{I}$, i.e., $\mathcal{T}_\mathrm{I}[\rho_\pm] = \rho_\mp$, leads to $b=c=0$ and thus
\begin{equation}
    \rho_\pm = a\,\rho_{++} \pm d\,\rho_{-+},
\end{equation}
which, with the unit trace condition ($a=1$) and the orthogonality condition ($d = \pm 1$), simply gives
\begin{equation}
    \rho_\pm = \rho_{++} \pm \rho_{-+}.
    \label{rho_pm_T1}
\end{equation}
We emphasize that, according to this, requiring the breaking of $\mathcal{T}_\mathrm{I}$ automatically implies that $\mathcal{T}_\mathrm{III}$ is preserved: $\mathcal{T}_\mathrm{III}[\rho_\pm] = \rho_\pm$. 
For the breaking of $\mathcal{T}_\mathrm{III}$, a similar calculation gives  
\begin{equation}
    \rho_\pm = \rho_{++} \pm \rho_{+-},
\end{equation}
which preserves $\mathcal{T}_\mathrm{I}$. We refer to these two ways of breaking $\mathcal{U}$ as \textit{directional spontaneous symmetry breaking}, and one can already note that it is reminiscent of $\mathcal{P}\mathcal{T}$ symmetry for open quantum systems as defined in Ref.~\cite{Huber2020}.

As is already apparent from Eq.~\eqref{A_alpha}, there are exactly two directions in which the $\mathbb{Z}_2$-symmetry can be broken, preventing the existence of a fourth phase. Indeed, requiring  $\mathcal{T}_\mathrm{I}[\rho_\pm] = \rho_\mp$ and $\mathcal{T}_\mathrm{III}[\rho_\pm] = \rho_\mp$ yields 
\begin{equation}
    \rho_+ = \rho_- = \rho_{++},
\end{equation}
so that the orthogonality condition can obviously no longer be satisfied. In this case, the steady state is unique and no SSB takes place as there is no mixing between the different symmetry sectors. Consequently, we have exhausted all possible directions to break the $\mathbb{Z}_2$ symmetry: either the system chooses to break $\mathcal{T}_\mathrm{I}$ or to break $\mathcal{T}_\mathrm{III}$, but never both at the same time; there is no ``fourth'' phase that would break both $\mathcal{T}_i$. Finally, we note that DSSB also explains why we end up with four different fixed points [see Eq.~\eqref{MF-phaseI} and Eq.~\eqref{MF-phaseIII}] from the breaking of a unique $\mathbb{Z}_2$-symmetry.

We emphasize that the concept of DSSB is not limited to our model, but is a general concept that can be applied to other multicritical systems. For instance, in Ref.~\cite{Soriente2018D} the authors discuss a $\mathbb{Z}_2 \times \mathbb{Z}_2$ symmetry that is a symmetry of the Hamiltonian~\cite{Baksic2014C}. However, once dissipation is incorporated, the model exhibits only a weak $\mathbb{Z}_2$ symmetry which is represented by a superoperator $\mathcal{U}$ that satisfies our decomposition~(\ref{Udecomp}), which explains the emergence of directional broken phases in their phase diagram.

In summary, we have identified a general structure that ensures DSSB:~the factorization of a unitary $\mathbb{Z}_2$-symmetry into two commuting antiunitaries $\mathcal{T}_\mathrm{I}$, $\mathcal{T}_\mathrm{III}$ that do not commute with the Liouvillian $\mathcal{L}_M$. In this case, the phases that spontaneously break the weak symmetry can be further distinguished according to the antiunitary they break and the one they preserve, which leads to different routes of spontaneous symmetry breaking.

\subsection{Finite-size results}

Now that the phases of our model and their breaking in the thermodynamic limit $N \to \infty$ has been established, we discuss how criticality emerges upon gradually increasing $N$. In addition, although (directional) spontaneous symmetry breaking occurs only in the thermodynamic limit, we show how to extract its signatures from the low-lying part of the spectrum of the generator of the dynamics at finite $N$.

\subsubsection{HEOM Liouvillian}
\label{sec:HEOM}
All the finite-size results that we present in this work come from the HEOM-based method introduced in Refs.~\cite{Debecker2024S, Debecker2024C}. We briefly outline here how the method can be applied to the Hamiltonian~\eqref{H_SE} with the correlation function~\eqref{CF}. The main object of interest is the \emph{HEOM Liouvillian}, denoted $\mathcal{L}_\mathrm{HEOM}$, which governs the numerically exact non-Markovian dynamics of the system (i.e., the spin degrees of freedom) for a factorized initial system-bath state. This dynamics is captured by an infinite hierarchy of equations that takes the following form~\cite{Tanimura89, Link2022}
\begin{align}
    \frac{d\rho^{({n}, {m})}}{dt} = &-i[H_{\mathrm{S}}, \rho^{({n}, {m})}] - [(n-m)i\omega+(n+m)\kappa]\rho^{({n}, {m})}\nonumber \\
    &+ G n S_x \rho^{({n}-1, {m})} + G m \rho^{({n}, {m}-1)}S_x \nonumber \\
      &+ [\rho^{({n}+1, {m})}, S_x] + [S_x, \rho^{({n}, {m}+1)}] ,
 \label{HEOM}
\end{align}
where $G = \gamma \kappa/2N$, $n, m \in \mathbb{N}$, $\rho^{(0, 0)} \equiv \rho_S$ and all the other $\rho^{(n, m)}$ are auxiliary operators necessary to capture the correct dynamics of the system. After truncation, Eq.~\eqref{HEOM} can be recast in the form
\begin{equation}
\frac{d}{dt}\dket{\rho} = \mathcal{L}_\mathrm{HEOM}(k_\mathrm{max}) \dket{\rho},
\label{HEOM_Liouvillian)}
\end{equation}
where $\dket{\rho}$ is a stacked vector containing all the vectorized versions of the operators $\rho^{(n, m)}$ under the usual isomorphism $\ket a \bra b \cong \ket a \ket b $. The operator $\mathcal{L}_\mathrm{HEOM}(k_\mathrm{max})$ defines the HEOM Liouvillian at a certain truncation order $k_\mathrm{max}$. This is because in practice the infinite hierarchy must be truncated. Here, we adopt a triangular truncation condition, where $\rho^{(n, m)} = 0$ for $n+m > k_\mathrm{max}$. It has been shown that $\mathcal{L}_\mathrm{HEOM}(k_\mathrm{max})$ provides a systematic and well-adapted pathway to study the spectral signatures of DPTs in the non-Markovian regime~\cite{Debecker2024S, Debecker2024C}. In the following section, we follow this approach.

\subsubsection{Emergence of criticality}

The emergence of two consecutive second-order DPTs as $N$ increases was previously examined in~\cite{Debecker2024C}. We therefore focus here on the first-order DPT, which describes the transition from phase (I) to phase (III). This phase transition can be seen as the merging of two critical points associated with the two consecutive second-order DPTs that separate phases (I) and (II) and phases (II) and (III), see Fig.~\ref{fig:PhaseDiagrams}. 

\begin{figure}
    \centering
    \includegraphics{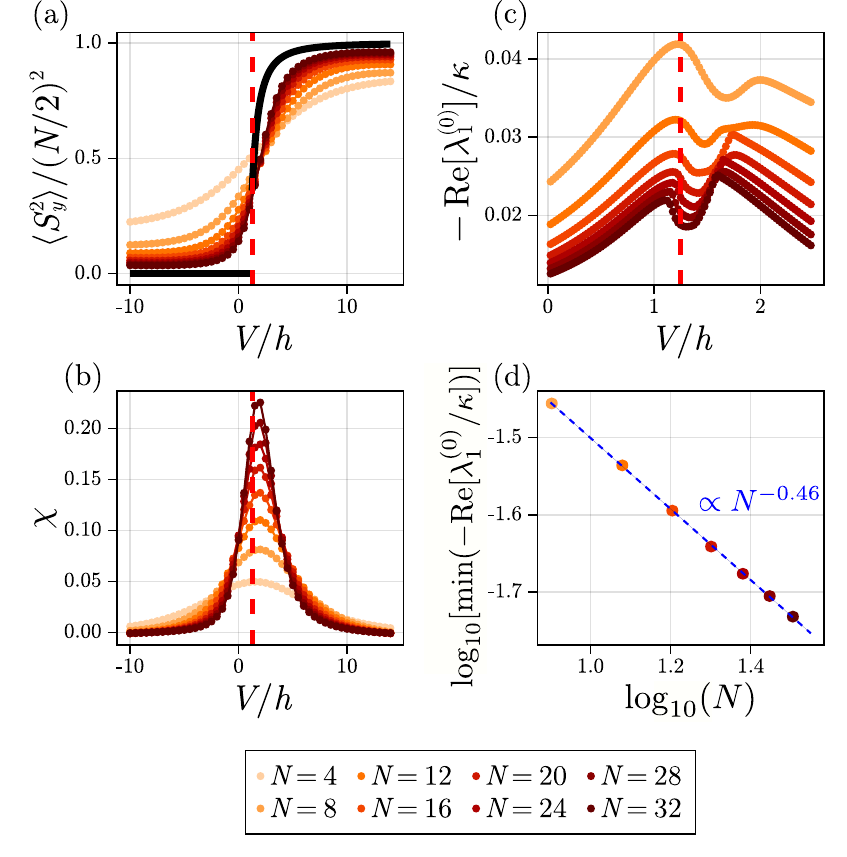}
    \caption{\textbf{Emergence of the first-order DPT.} (a): Mean value of $\braket{S_y^2}/(N/2)^2$ as a function of the control parameter $V/h$ showing the emergence of a first-order DPT as we approach the thermodynamic limit $N\rightarrow +\infty$. The straight black line displays the exact result (mean-field) in the thermodynamic limit.  (b): Susceptibility $\chi$ [Eq.~(\ref{susceptibility})] signaling the emergence of a discontinuity at the critical point $V/h = q_2\gamma/(4h)$, represented by a dashed red line. (c): Gap in the symmetry sector $k=0$ showing the emergence of a local minimum near the critical point. In the thermodynamic limit, the gap vanishes at the critical point as suggested by the finite-size scaling of the local minimum with respect to $V/h$ depicted in panel (d). The parameters are $\kappa = \omega  = 10 h$, $\gamma = 5 h/q_2$, and $k_\mathrm{max} = 11$ for panels (a), (b) and $k_\mathrm{max} = 17$ for panels (c), (d).}
    \label{Fig_sy2}
\end{figure}

Phase (I) is characterized by $\braket{S_y^2} = 0$, while phase (III) is marked by $\braket{S_y^2} \neq 0$. Consequently, we may define the normalized quantity $\braket{S_y^2}/(N/2)^2$ as a possible order parameter of the phase transition. In Fig.~\ref{Fig_sy2}~(a), we illustrate the behavior of this order parameter for various system sizes. The transition becomes sharper as $N$ grows, and in the thermodynamic limit $N \rightarrow +\infty$, the exact mean-field prediction (black curves) exhibits a discontinuity at the critical point. While finite-$N$ curves will always be continuous, we can explore the onset of discontinuity by examining the behavior of the susceptibility $\chi$ that we define by
\begin{equation}\label{susceptibility}
  \chi \equiv \frac{d}{dv}\left[\frac{\braket{S_y^2}}{(N/2)^2}\right], \quad v \equiv V/h.
\end{equation}
Figure~\ref{Fig_sy2}~(b) shows that, as expected, $\chi$ reaches a maximum close to the critical point (indicated by the red dashed line), and that the height of the maximum increases with $N$ while its location approaches the critical point.

A defining hallmark of a first-order DPT is the emergence -- in the thermodynamic limit and in the symmetry sector $k=0$ containing the steady state -- of an eigenvalue $\lambda_1^{(0)}$ whose real part vanishes only at the critical point, while remaining nonzero in its finite vicinity~\cite{Debecker2024S, Minganti2018Spectral}. In other words, the gap in the sector $k=0$ must vanish exactly at the critical point. We confirm this behavior via finite-size numerics, as shown in Fig.~\ref{Fig_sy2}~(c). A local minimum in $-\mathrm{Re}[\lambda_1^{(0)}]$ deepens with $N$ and shifts towards the critical point. Figure~\ref{Fig_sy2}~(d) further indicates that the gap closes with a power-law scaling at the critical point.

Interestingly, the scaling of the gap has recently been connected to phase coexistence~\cite{Ptaszy2024D}. In agreement with the results of Ref.~\cite{Ptaszy2024D}, we find that the power-law scaling could have been guessed from the existence of the infinite number of attractors emerging at the critical point, as discussed in Subsec.~\ref{section::MF}. Finally, we note that the imaginary part of $\lambda_1^{(0)}$ should also vanish around the critical point~\cite{Minganti2018Spectral, Debecker2024S}. Here, we find that over the range of parameters considered, $\mathrm{Im}[\lambda_1^{(0)}]$ is identically zero (data not shown).

\begin{figure*}
    \centering
    \includegraphics[width = \linewidth]{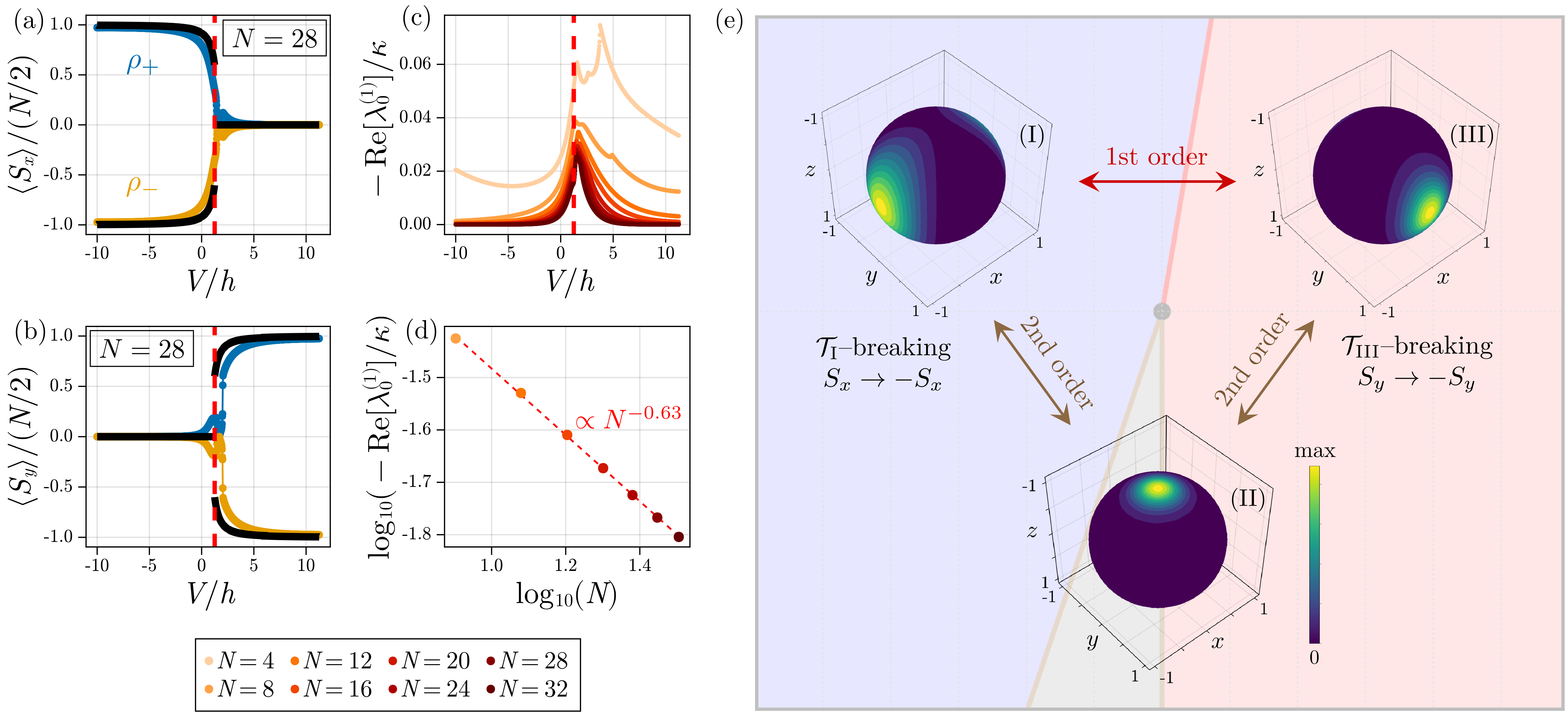}
    \caption{\textbf{Emergence of DSSB.} (a) and (b): Expectation values $\braket{S_x}$ and $\braket{S_y}$ as a function of the control parameter $V/h$ for the states $\rho_\pm$ directly extracted from the HEOM Liouvillian for $N = 28$ [see Eq.~\eqref{eq::rhopm}]. The straight black line displays the exact result (mean-field) in the thermodynamic limit and the red dashed line indicates the critical point. We observe a good agreement between the finite-size curves and the mean-field results, which confirms the DSSB mechanism.  (c): Gap in the symmetry sector $k=1$ as a function of $V/h$ indicating a gap closing in phase (I) and (III). (d): Scaling of the gap at the critical point showing a power law gap closing. The parameters are $\kappa = \omega  = 10 h$, $\gamma = 5 h/q_2$ and $k_\mathrm{max} = 10 $. (e) Density plots of the Husimi function on the sphere for the three distinct phases. These plots show how the system transitions between phases: from phase I to II and from phase II to III via continuous (second-order) transitions, where the Husimi function smoothly deforms through the symmetric phase II. In contrast, the transition from phase I to III is discontinuous (first-order), marked by squeezing in perpendicular directions. The parameters used are $N=45$, $\kappa = 0.2\gamma$, $\omega = 0.5\gamma$, $k_{\mathrm{max}}=7$ and $\gamma q_2/(4h) \approx 0.086$ and $V/h = 0$ for phase (II), and $\gamma q_2/(4h) \approx 1.724$ and $V/h = \pm 50$ for phases (I) and (III).}
    \label{fig:DSSB}
\end{figure*}

\subsubsection{Emergence of Directional Spontaneous Symmetry Breaking}

Strictly speaking, SSB in open quantum systems only arises in the thermodynamic limit. This also applies a fortiori for the DSSB scenario discussed above. In order to investigate the emergence of DSSB at finite $N$, we numerically compute [see Eq.~\eqref{rho_pm}]
\begin{equation}
    \rho_0^{(1)} \propto \rho_+ -  \rho_-
    \label{eq::rhopm}
\end{equation}
According to the general theory in Subsec.~\ref{subsubsec:fromSSBtoDSSB}, the states $\rho_\pm$ should capture the breaking of $\mathcal{T}_\mathrm{I}$ and the preservation of  $\mathcal{T}_\mathrm{III}$ in phase (I), and vice versa in phase (III). This behavior is evident in Fig.~\ref{fig:DSSB}~(a)-(b) where $\mathcal{T}_\mathrm{I}$-breaking is signaled by two branches (associated with $\rho_\pm$) that emerge in the expectation value of $S_x$, while $\mathcal{T}_\mathrm{III}$-breaking is indicated by two branches in $\braket{S_y}$~\footnote{Recall that the finite-size results displayed in Fig.~\ref{fig:DSSB} are obtained via the HEOM Liouvillian~(\ref{HEOM}), rather than from the Markovian embedding [Eq.~\eqref{master_LMG_tot}] which includes the pseudomode in the system. As a consequence, the solution includes auxiliary operators not directly relevant here, and one must select the $(0, 0)$ component of the appropriate eigenvector, as covered in detail in Ref.~\cite{Debecker2024S}}. In addition, these finite-$N$ curves closely follow the mean-field predictions shown by the black curves in the same panels. Apart from minor finite-size deviations around the critical point (red dashed line), it is clear that the system can break $\mathcal{T}_\mathrm{I}$ or $\mathcal{T}_\mathrm{III}$ but \emph{not} both simultaneously. 

In addition, in Fig.~\ref{fig:DSSB}~(c), we plot the Liouvillian gap $- \mathrm{Re}[\lambda_0^{(1)}]$ in the $u_{k=1} = -1$ sector as a function of the control parameter $V/h$ for different values of $N$. Because phases (I) and (III) spontaneously break $\mathcal{U}$, we expect this gap to close in the thermodynamic limit in those broken symmetry phases. At the critical point, phases (I) and (III) meet and mean-field theory predicts an infinite continuum of steady states - akin to a $\mathbb{U}(1)$-symmetry breaking - hence the gap must also vanish there in the thermodynamic limit. Fig.~\ref{fig:DSSB}~(d) confirms this by showing that the gap closes as $N^{-0.63}$ at the critical point. 

Finally, in Fig.~\ref{fig:DSSB}~(e) , we show the Husimi functions of the steady states in each phase. Since, in the broken-symmetry phases, the steady states $\rho_{\mathrm{ss}}$ can be written in terms of the symmetry-broken states $\rho_{\pm}$ as~\cite{Minganti2018Spectral, Debecker2024S}
\begin{equation}
    \rho_{\mathrm{ss}} \approx \frac{\rho_+ + \rho_-}{2},
    \label{eq::asymptotic_rhoss}
\end{equation}
for $N$ large enough, we expect finite-size consequences of the DSSB on the steady states. Indeed, in phase (I), we know that $\mathcal{T}_\mathrm{I}$ is broken, but not $\mathcal{T}_\mathrm{III}$, which directly translates in the two branches in $\braket{S_x}$ in Fig.~\ref{fig:DSSB}~(a) while $\braket{S_y} = 0$ as shown in panel (b). Consequently, since Eq.~\eqref{eq::asymptotic_rhoss} is a asymptotic good approximation of $\rho_{\mathrm{ss}}$, we expect the Husimi function to be composed of two peaks located at $x = +1$ and $x = -1$ ($\braket{S_x} = 0$ for every finite $N$), while it should be close to zero elsewhere and, in particular, in the $y$ direction. Of course, in the strict thermodynamic limit $N \rightarrow +\infty$, the system would break $\mathcal{T}_I$ (and $\mathcal{U}$) by choosing only one of the two peaks. Similar considerations hold for phase (III), with the appropriate replacement $\mathcal{T}_\mathrm{I} \rightarrow \mathcal{T}_\mathrm{III}$ and $x \rightarrow y$. These finite-size implications are clearly visible in Fig.~\ref{fig:DSSB}~(e) where two peaks appear along the $\pm x$ and $\pm y$ directions for phases (I) and (III), respectively. In phase (II), the Husimi function is concentrated around $z=-1$, as expected from mean-field theory~\eqref{MF-phaseII}. Furthermore, the Husimi function exhibits squeezing in the $xy$-plane in the direction orthogonal to the direction of symmetry breaking: squeezing along $y$ and symmetry breaking along $x$ in phase (I) and squeezing along $x$ and symmetry breaking along $y$ in phase (III). Hence, spin squeezing, on which we elaborate in the next section, serves in our model as an explicit indicator (and physical signature) of the emergence of a preferred direction for breaking the symmetry, i.e., of directional symmetry breaking.

\section{Spin squeezing}
So far, our analysis has primarily focused on comparing finite-size results with mean-field predictions. However, as illustrated in Fig.~\ref{fig:DSSB}~(e) and discussed in the previous section, the model is richer than the mean-field theory. This motivates the investigation of fluctuations around the mean-field predictions, which we undertake by examining the behavior of the spin squeezing.

Spin squeezing is a key indicator of the non-classical nature of spin states, and directly quantifies the gain achievable in quantum metrology to exceed classical limits of precision. Importantly, spin squeezing can be directly measured in the laboratory through collective spin measurements, making it a practical tool. Here, we study the spin squeezing properties of the steady state, both for the thermodynamic limit and for finite $N$. 
 We show that spin squeezing occurs near the critical points, in a region whose extent is determined by the ratios $h/\omega$ and $\kappa/\omega$. In other words, phase transitions may or may not be accompanied by spin squeezing, depending on the specific choice of parameters. Furthermore, we show that squeezing becomes more pronounced with increasing bath memory (i.e., decreasing $\kappa$) and with increasing magnetic field strength $h$. 

\subsection{Spin squeezing parameter}

To characterize the amount of spin squeezing carried by the steady state, we use the spin squeezing parameter introduced by Kitagawa and Ueda~\cite{Kitagawa1993S}
\begin{equation}
   \xi^2 = \frac{4~\mathrm{min}_{\vec{n}_\perp} (\Delta S_{\vec{n}_\perp})^2 }{N},
    \label{Ueda}
\end{equation}
with the minimum of the variance computed with respect to all unit directions $\mathbf{n}_\perp$ orthogonal to the mean spin direction $\braket{\mathbf{S}}$. A state is said to be spin-squeezed if $\xi^2 <1$ and it can be shown that the symmetric collective spin-squeezed states are pairwise entangled~\cite{Korbicz2005S, Pezze2018Q}. For a mean spin direction along the $z$ axis, one can show~\cite{Ma2011Q} that $\xi^2$ reduces to

\begin{equation}\label{spinsqueezingalongz}
    \xi^2 = \frac{2}{N}\left[\braket{S_{x}^2 + S_{y}^2} - \sqrt{\braket{S_{x}^2 - S_{y}^2}^2 + \braket{\{S_{x},S_{y}\}}^2}\right].
\end{equation}
Note that for the LMG model~\eqref{master_LMG_tot} considered here and for \textit{any finite $N$}, the mean spin direction will always be the $z$ direction as the $\mathbb{Z}_2$ symmetry directly implies that $\braket{S_x} = \braket{S_y} = 0$. However, in the thermodynamic limit $N \rightarrow +\infty$, the symmetry $\mathbb{Z}_2$ is explicitly broken in phases (I) and (III), which prevents such identifications  [see Eqs.~\eqref{MF-phaseI} and \eqref{MF-phaseIII}].

\subsection{Analytical derivation by third quantization}

Computing spin squeezing requires going beyond the mean-field results presented in Section~III, as we need to access quantum fluctuations around the mean-field solutions. Here, we outline how one can exactly solve the LMG model~\eqref{master_LMG_tot} in the thermodynamic limit in phase~(II). 

We start by representing the spin degrees of freedom by bosons through the Holstein-Primakoff (HP) transformation
\begin{equation}
    \begin{split}
        S_+ &= \sqrt{N}b^\dagger \sqrt{1 - b^\dagger b/N},\\
        S_- &= \sqrt{N}\sqrt{1 - b^\dagger b/N}~b, \\
        S_z &= b^\dagger b - \frac{N}{2},
    \end{split}
    \label{HP}
\end{equation}
with $b$ ($b^\dagger$) the annihilation (creation) operator of an effective bosonic mode. In phase (II), mean-field calculations gave $\langle S_z\rangle_\mathrm{ss} = -N/2$ [see Eq.~\eqref{MF-phaseII}] and thus fluctuations satisfy $\braket{b^\dagger b} \ll N$. Expanding the argument of the square roots to zeroth order ($\tiny \sqrt{1-b^\dagger b/N} \approx 1$) leads for the Hamiltonian \eqref{H_LMG_tot} to the effective Hamiltonian
\begin{equation}
   H_\mathrm{HP}^{\mathrm{(II)}} = \frac{V}{2}(b^{\dagger 2} + b^2) + h b^\dagger b + \omega a^\dagger a + \frac12 \sqrt{\frac{\gamma \kappa}{2}} (b+b^\dagger) (a+a^\dagger),
   \label{HLMGII}
\end{equation}
up to $1/\sqrt{N}$ corrections. In the thermodynamic limit, the dynamics of the enlarged Markovian system made of the spins and the pseudomode becomes exactly described by the Liouvillian
\begin{equation}
   \mathcal{L}_\mathrm{HP}^{\mathrm{(II)}}[\bigcdot] = -i\big[H_\mathrm{HP}^{\mathrm{(II)}}, \bigcdot\big] + \kappa (2a \bigcdot a^\dagger - \{a^\dagger a, \bigcdot\}).
   \label{master_LMG_HP}
\end{equation}
Being quadratic in the bosonic operators $a$ and $b$, the superoperator $\mathcal{L}_\mathrm{HP}^{\mathrm{(II)}}$ can be diagonalized by the third quantization method~\cite{Prosen2008T, Prosen2010Q}. In particular, this readily gives the expectation values of all second-order moments, from which the spin squeezing parameter can be evaluated through Eq.~(\ref{spinsqueezingalongz}).

Expanding the square root in the transformation \eqref{HP} is a priori not justified in phases (I) and (III) because the condition $\braket{b^\dagger b} \ll N$ is not met. This apparent problem can be circumvented by first doing a rotation of the spin degrees of freedom [along with a shift of the pseudomode for phase (I)] to ensure that the mean spin direction is along the $z$ axis, and then applying the HP transformation~\eqref{HP}, as explained in detail in Appendix~\ref{Sec:AppTQ} based on the works~\cite{Ma2009F, Dusuel2005C, Emary2003C, Buca2019D, Morrison2008C, Fan2023C}. Using this, we are therefore able to obtain the exact expressions of the spin squeezing in the thermodynamic limit in all regimes of parameters. \\

\subsection{Dissipative phase transitions and spin squeezing}
\label{subsec::DPTandSpinSqueezing}

We now discuss the connections between spin squeezing and dissipative phase transitions in our model.

\subsubsection{$2^\mathrm{nd}$-order DPTs and spin squeezing}
\begin{figure*}
    \centering
\includegraphics{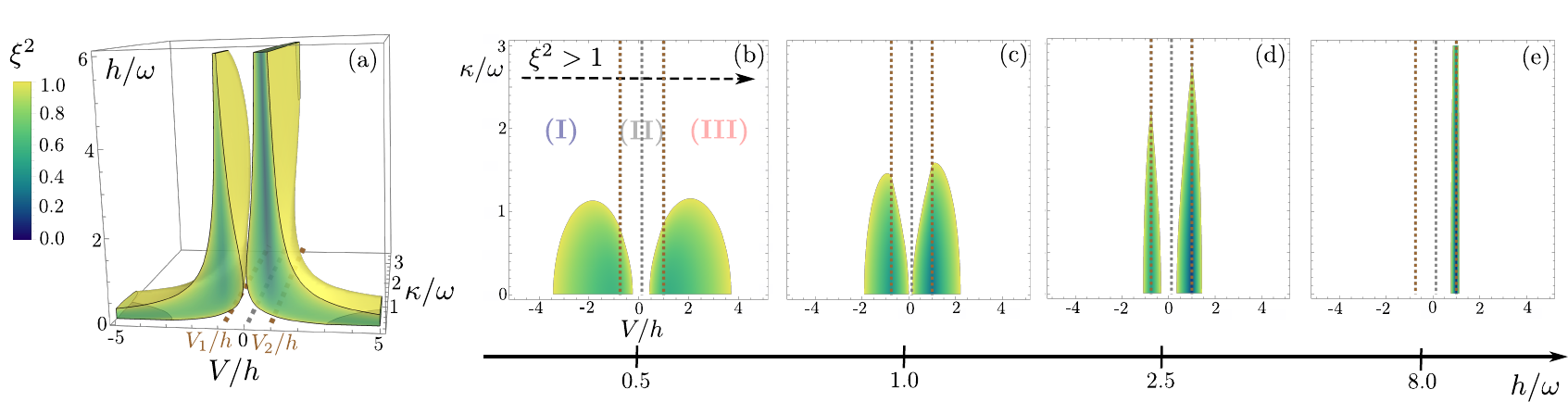}
    \caption{\textbf{Spin squeezing across 2$^{\mathrm{nd}}$-order DPTs.}
    (a): Exact spin squeezing parameter $\xi^2$ in the thermodynamic limit as a function of $V/h$, $h/\omega$ and $\kappa/\omega$ for $\gamma = h/(2q_2)$, highlighting the squeezed regions (i.e., regions where $\xi^2 <1$). The brown dashed line indicates the critical points.  (b)--(e): Density plots of the squeezing parameter as a function of the parameters $V/h$ and $\kappa/\omega$ for fixed values of $h/\omega$, corresponding to different horizontal slices of the 3D plot shown in panel (a). Decreasing $\kappa/\omega$ (i.e., increasing memory effects) allows for a stronger squeezing, i.e., a decrease of $\xi^2$. The gray dashed line indicates the special line $V = q_2\gamma/4$ where no spin squeezing is possible (see main text). All panels show that the DPTs may or may not be accompanied by spin squeezing: a path in parameter space that connects the three different phases without involving spin squeezing is highlighted in panel (b) by a black dashed arrow.}
    \label{fig:ParameterSpace3DSS}
\end{figure*}

\begin{figure}
    \centering
\includegraphics{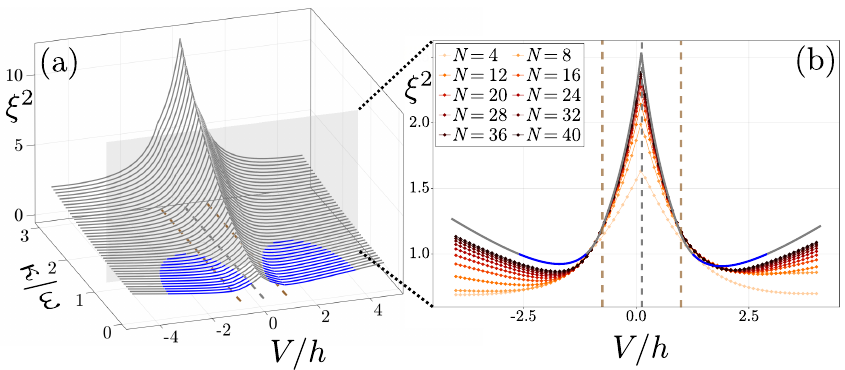}
    \caption{\textbf{Spin squeezing and finite-size effects.} (a): Exact spin squeezing parameter $\xi^2$ in the thermodynamic limit (gray curves) as a function of  $V/h$ and $\kappa/\omega$ for $\gamma = h/(2q_2)$ and $h=\omega/2$. The blue part of the curves corresponds to the squeezed region (where $\xi^2 < 1$). The two brown dashed lines indicate the two critical points. When $\kappa/\omega$ decreases, a squeezing region appears within the three phases, which shows that continuous phase transitions can occur with or without spin squeezing. (b): For $\kappa/\omega = 1.0$ [gray plane in (a)], comparison of the exact result in the thermodynamic limit with the finite-size results for different values of $N$ (see legend) with $k_\mathrm{max} = 8$. The squeezing parameter is maximum for $V/h= (V_2+V_1)/2h$, indicated by the thin gray dashed line.}
    \label{fig:SpinS_finiteNvsThermo}
\end{figure}

\begin{figure}
\includegraphics{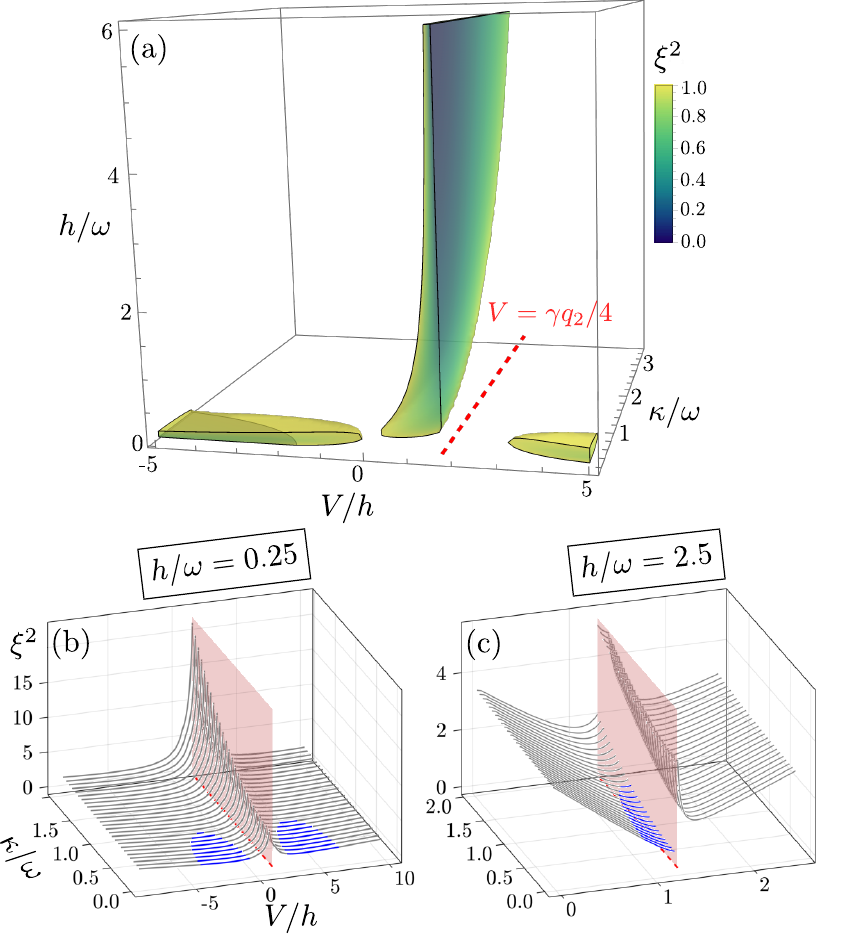}
   \caption{\textbf{Spin squeezing across 1$^{\mathrm{st}}$-order DPT}. (a): Exact spin squeezing parameter $\xi^2$ in the thermodynamic limit as a function of $V/h$, $h/\omega$ and $\kappa/\omega$ for $\gamma = 5h/q_2$, highlighting the squeezed regions (i.e., regions where $\xi^2 <1$). (b): Squeezing parameter as a function of $V/h$ and $\kappa/\omega$ for $h/\omega = 1/4$, showing that it can vary either continuously or discontinuously with the control parameter $V/h$. The blue region indicates values where $\xi^2 <1$. (c): same as (b) for $h/\omega = 5/2$. The red plane (line for panel~(a) to ensure clarity) is the critical plane separating phase (I) from phase (III).}
    \label{fig:Squeezing1stOrder}
\end{figure}

When $\gamma q_2/(4h) < 1$, the three different phases are separated by two second-order DPTs, as shown in Fig.~\ref{fig:PhaseDiagrams}~(a). To explore the behavior of the spin squeezing parameter, we focus on a fixed horizontal line within the phase diagram. Consequently, in the following analysis, we set $\gamma = h/(2 q_2)$, although similar results are obtained for fixed values of $\gamma$ satisfying $\gamma  < 4h/ q_2$.

In Fig.~\ref{fig:ParameterSpace3DSS}~(a), we present the region in the parameter space where spin squeezing occurs in the thermodynamic limit ($N \rightarrow +\infty$), i.e., when $\xi^2 < 1$. The results reveal that spin squeezing emerges near the critical points, in a region whose size depends on the structure of the bath captured by $\kappa/\omega$. Notably, for a fixed value of $h/\omega$, increasing the bath memory (i.e., decreasing $\kappa$) decreases the spin squeezing parameter, as shown in Fig.~\ref{fig:ParameterSpace3DSS}~(b-e) where the complementary role of the magnetic field in determining spin squeezing is also evident. Increasing $h/\omega$ makes it possible to obtain squeezing in narrower regions of parameters around the critical points along the $V/h$ axis, but in wider regions along the $\kappa/\omega$ (‘memory’) axis. This shows that memory effects can protect the quantum fluctuations necessary for spin squeezing.\\

Remarkably, although spin squeezing is, when present, roughly more pronounced around the critical points, the presence of spin squeezing is not inherently tied to the phase transitions; it may or may not accompany them depending on the system parameters. This indicates that spin squeezing is not a defining characteristic of any specific phase. Within phase (II), we also observe that nearly every point in the parameter space can exhibit spin squeezing [see Fig.~\ref{fig:ParameterSpace3DSS}~(a) and (c)]. The only exception is the plane defined by points equidistant from both critical points, corresponding to $V = q_2 \gamma/4$. As discussed in Section~III, at this special value of $V$, the steady state becomes $\mathbb{U}(1)$-symmetric in the thermodynamic limit, resulting in a steady state that must be diagonal in the $S_z$ basis and, consequently, incapable of supporting spin squeezing. 

This behavior highlights the observable consequences of the $\mathbb{U}(1)$-symmetry that emerges in the thermodynamic limit for $V = q_2 \gamma/4$. In fact, the spin squeezing parameter hits a local maximum larger than $1$ for $V = q_2 \gamma/4$, as shown in Fig.~\ref{fig:SpinS_finiteNvsThermo}~(a). In Fig.~\ref{fig:SpinS_finiteNvsThermo}~(b), we compare finite-size results and the exact results in the thermodynamic limit. As expected, finite-size curves mainly deviate from the analytical predictions in the thermodynamic limit around the critical points.

\subsubsection{$1^\mathrm{st}$-order DPT and spin squeezing} 
After having analyzed the behavior of the spin-squeezing parameter across the three different phases separated by second-order DPTs, a natural question arises: how does the spin-squeezing parameter behave when phases (I) and (III) are connected discontinuously in the thermodynamic limit ? To address this, we set $\gamma = 5h/q_2$, which ensures that phases (I) and (III) are connected by a first-order DPT (see Fig.~\ref{fig:PhaseDiagrams}), and we study the spin squeezing parameter as a function of $V/h,~~\kappa/\omega$ and $h/\omega$. 

In Fig.~\ref{fig:Squeezing1stOrder}~(a), we delineate the parameter regions where spin squeezing is observed. In particular, as in the case of 2$^{\mathrm{nd}}$-order DPTs (Fig.~\ref{fig:ParameterSpace3DSS}), spin squeezing occurs when the bath has sufficient memory, quantified by a sufficiently small $\kappa/\omega$ ratio. However, we observe that the first-order DPT is generally associated with a less significant spin squeezing. The spin-squeezing regions fragment into three distinct domains, which give rise to different kinds of behavior.

For weak magnetic fields, spin squeezing occurs both before and after the critical point, provided that $\kappa/\omega$ remains small. In this regime, the spin squeezing parameter evolves continuously with the control parameter $V/h$, reaching a maximum at the critical point, as shown in  Fig.~\ref{fig:Squeezing1stOrder}~(b) for $h/\omega = 0.25$. In contrast, in the strong magnetic field regime, spin squeezing is restricted to a narrow region preceding the critical point (again for a sufficiently small region $\kappa/\omega$), while the state immediately beyond the critical point is never squeezed. This results in a discontinuity of the squeezing parameter at the critical point, as illustrated in Fig.~\ref{fig:Squeezing1stOrder}~(c).  

In summary, the first-order DPT may occur with or without spin squeezing, and the critical point can manifest itself either as a maximum or as a point of discontinuity for the spin squeezing parameter, underscoring the complex interplay between criticality and quantum correlations in the non-Markovian regime.

\section{Proposed experimental implementation}
\label{section:exp}

\begin{figure}
	\includegraphics[width=0.65\linewidth]{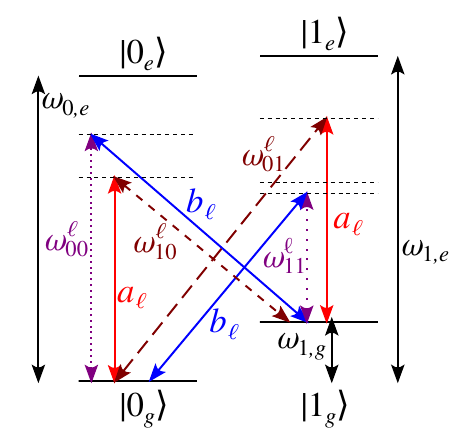}
	\caption{\textbf{Experimental proposal.} Proposed level scheme, together with driving fields at frequencies $\omega_{jj'}^\ell$ (dotted and dashed lines) and cavity modes $a_\ell$, $b_\ell$ (solid lines), where $j,j'=0,1$, $\ell=1,2$. For ease of visualization, only one $\ell$ is shown. Notations are explained in the main text.}
	\label{fig:LevelScheme}
\end{figure}

Let us finally discuss a possible experimental implementation of our LMG model in the framework of cavity QED, following \cite{Dimer2007Jan,Morrison2008,Morrison2008C} (compare also \cite{Pausch2024}). We consider $N$ atoms with a ground-state manifold of two states $\ket{0_g}$, $\ket{1_g}$ at energies $\omega_{0,g} = 0$, $\omega_{1,g}$ and an excited-state manifold of two states $\ket{0_e}$, $\ket{1_e}$ at energies $\omega_{0,e}$, $\omega_{1,e}$ ($\hbar = 1$ for simplicity). As sketched in Fig.~\ref{fig:LevelScheme}, the transitions between these manifolds are driven off-resonantly by eight lasers of Rabi frequencies $\Omega_{jj'}^\ell$ and driving frequencies $\omega_{jj'}^\ell$ ($\ell=1,2$, $j,j'=0,1$) and they couple off-resonantly to four cavity modes $a_1, a_2, b_1, b_2$. The polarizations of the driving fields and of the modes are chosen such that $a_\ell$ ($b_\ell$), with $\ell=1,2$, couples only to the $\ket{j_g}\leftrightarrow\ket{j_e}$ transitions (only to the $\ket{j_g}\leftrightarrow\ket{(1-j)_e}$ transitions) and $\Omega_{jj'}^\ell$ only to the transitions where the index changes by $j'-j$ from ground state to excited state.
The coupling strengths of the atoms and the cavity modes $a_\ell$, $b_\ell$ are $g_{jj'}^\ell$ for the cavity-mediated transition $\ket{j_g}\leftrightarrow\ket{j'_e}$.
Note that in fact only one of the modes $a_2$, $b_2$ with its corresponding driving fields $(\Omega_{10}^2,\Omega_{01}^2)$, $(\Omega_{00}^2,\Omega_{11}^2)$ is needed to prepare our model.

We assume that the frequencies fulfill the following resonance conditions (compare Fig.~\ref{fig:LevelScheme}):
\begin{subequations}
\begin{align}
	\omega_{a_{\ell}} &\approx \omega_{10}^\ell + \omega_{1,g} \approx \omega_{01}^\ell - \omega_{1,g}, \\ \omega_{b_{\ell}} &\approx \omega_{11}^\ell + \omega_{1,g} \approx \omega_{00}^\ell - \omega_{1,g},
\end{align}
and consequently also
\begin{align}
	\omega_{01}^\ell - \omega_{10}^\ell &\approx 2 \omega_{1,g}, & \omega_{00}^\ell - \omega_{11}^\ell &\approx 2 \omega_{1,g},
\end{align}
\end{subequations}
while other combinations of frequencies are assumed to be off-resonant, i.e., their absolute difference (such as $|\omega_{11}^\ell-\omega_{a_{\ell}}|$) is sufficiently large and much larger than the absolute difference of the resonant frequencies (such as $|\omega_{a_{\ell}} - \omega_{10}^\ell - \omega_{1,g}|$).

We first transform into the interaction picture with
\begin{align}
	\nonumber
	H_0 ={}& \sum_{i=1}^{N} \left[ \omega_{1,g}' \ket{1_g}_i\bra{1_g}_i + \sum_{j=0,1} \omega_{0j}^1 \ket{e_j}_i\bra{e_j}_i  \right]\\
	&+ \sum_{k=a_1,a_2, \atop b_1,b_2} \omega_k' k^\dagger k,
\end{align}
where $\omega_k'$ and $\omega_{1,g}'$ are frequencies close to (or identical to) $\omega_k$ and $\omega_{1,g}$. In particular, we define $\omega_{a_{\ell}}' = \omega_{01}^\ell - \omega_{1,g}'$, $\omega_{b_{\ell}}' = \omega_{00}^\ell -\omega_{1,g}'$ and we demand that $\omega_{01}^\ell - \omega_{10}^\ell = \omega_{00}^\ell - \omega_{11}^\ell = 2 \omega_{1,g}'$. Assuming that the frequency differences $\Delta_j = \omega_{j,e} - \omega_{0j}^1$ ($j=0,1$) 
are much larger than any other frequency scale characterizing the system, i.e., $\big|\Delta_j\big| \gg \big|\omega_k - \omega_k'\big|, \big|\omega_{1,g} - \omega_{1,g}'\big|, \big|g_{jj'}^\ell\big|, \big|\Omega_{jj'}^\ell\big|$, we can adiabatically eliminate the excited-state manifold~\cite{Dimer2007Jan,Morrison2008C}, yielding (after omission of global energy shifts) the effective ground-state Hamiltonian 
\begin{align}
	&H_\text{eff} = \omega_0  S_z  \nonumber\\
    &+ \sum_{k=a_1,a_2, \atop b_1,b_2} \left[\delta_k k^\dagger k - 2 \delta_k^- S_z k^\dagger k - \frac{\lambda_k}{\sqrt{N}}\left(X_k k + X_k^\dagger k^\dagger\right)\right],
\end{align}
with collective operators
\begin{subequations}
\begin{align}
	S_z &= \frac{1}{2} \sum_{i=1}^{N} \left(\ket{1_g}_i\bra{1_g}_i - \ket{0_g}_i\bra{0_g}_i\right), \\ S_+ &= \sum_{i=1}^{N} \ket{1_g}_i\bra{0_g}_i, \quad S_- = S_+^\dagger,
\end{align}
\end{subequations}
with $X_k = \alpha_k S_+ + \beta_k S_-$ ($k={a_1,a_2,b_1,b_2}$, dimensionless parameters $\alpha_k$, $\beta_k$) and with parameters~\footnote{Note that these parameters differ slightly from those presented in Ref.~\cite{Morrison2008C}, since there was a sign error in that publication and the authors of Ref.~\cite{Morrison2008C} omitted a contribution to $\omega_0$ that we are including.}
\begin{subequations}
\begin{align}
	\nonumber
	&\omega_0 ={} \omega_{1,g}-\omega_{1,g}' \\
	&- \frac{1}{4}\sum_{\ell=1,2} \left[\sum_{j'=0,1} \left(\frac{|\Omega_{j'j'}^\ell|^2}{\Delta_1} - \frac{|\Omega_{j'j'}^\ell|^2}{\Delta_0}\right) + \frac{|\Omega_{10}^\ell|^2}{\Delta_{0}} - \frac{|\Omega_{01}^\ell|^2}{\Delta_{1}}\right],\\
	&\delta_k ={} \left(\omega_k - \omega_k'\right) - N \delta_k^+,\\
	&\delta_{a_\ell}^\pm ={} \frac{1}{2}\left(\frac{|g_{11}^\ell|^2}{\Delta_1} \pm \frac{|g_{00}^\ell|^2}{\Delta_0}\right), \quad \delta_{b_\ell}^\pm = \left(\frac{|g_{10}^\ell|^2}{\Delta_{0}} \pm \frac{|g_{01}^\ell|^2}{\Delta_{1}} \right),\\
	&\frac{\lambda_{a_\ell} \alpha_{a_\ell}}{\sqrt{N}} ={} \frac{\left(\Omega_{10}^\ell\right)^*g_{00}^\ell}{2 \Delta_0}, \qquad \frac{\lambda_{a_\ell} \beta_{a_\ell}}{\sqrt{N}} = \frac{\left(\Omega_{01}^\ell\right)^* g_{11}^\ell}{2 \Delta_1},\\
	&\frac{\lambda_{b_\ell} \alpha_{b_\ell}}{\sqrt{N}} ={} \frac{\left(\Omega_{11}^\ell\right)^* g_{01}^\ell}{2\Delta_1}, \qquad \frac{\lambda_{b_\ell} \beta_{b_\ell}}{\sqrt{N}} = \frac{\left(\Omega_{00}^\ell\right)^* g_{10}^\ell}{2 \Delta_0}.
\end{align}
\end{subequations}
Note that the coupling strengths are given as $g_{jj'}^\ell = g_0^\ell c_{j,j'}$ with Clebsch-Gordan coefficients $c_{0,0} = c_{1,1}$, $c_{0,1} = c_{1,0}$ \cite{MetcalfBook}, such that $\delta_k^- = 0$ if $\Delta_0 = \Delta_1$.

We now proceed by assuming the cavity modes to be damped, i.e., the full dynamics is given by
\begin{align}
	\dot{\rho}_\text{full} = -\i \left[H_\text{eff},\rho_\text{full}\right] + \sum_{k=a_1,a_2, \atop b_1,b_2} \kappa_k \left(2 k \rho_\text{full} k^\dagger - \{ k^\dagger k, \rho_\text{full} \}\right),
\end{align}
where the dynamics of the modes labelled by $\ell=1$ is assumed to define the fastest time scale of the system, i.e., $\sqrt{\delta_{a_1}^2 + \kappa_{a_1}^2}, \sqrt{\delta_{b_1}^2 + \kappa_{b_2}^2} \gg \omega_0, \lambda_{k}, \delta_{a_2}, \delta_{b_2}, \kappa_{a_2}, \kappa_{b_2}$. We can then adiabatically eliminate also the modes $a_1$, $b_1$ \cite{Azouit2017}  and obtain
\begin{align}
	\nonumber
	\dot{\rho} ={}& -\i [H,\rho] + \sum_{k=a_2,b_2} \kappa_k\left(2 k \rho k^\dagger - \left\{ k^\dagger k, \rho\right\} \right) \\
	&+ \sum_{k=a_1,b_1} \frac{\lambda_{k}^2}{N} \frac{\kappa_k}{\kappa_k^2 + \delta_k^2} \left( 2   X_k^\dagger\rho X_k - \left\{X_k X_k^\dagger, \rho\right\}\right),
\end{align}
with 
\begin{align}
	\nonumber
	H ={}& \omega_0 S_z + \sum_{k={a_2,b_2}} \left[\left( \delta_k   - 2 \delta_k^- S_z \right) k^\dagger k - \frac{\lambda_k}{\sqrt{N}} \left(X_k k + X_k^\dagger k^\dagger\right) \right] \\
	&- \sum_{k={a_1,b_1}} \frac{\lambda_{k}^2}{N} \frac{\delta_k}{\kappa_k^2 + \delta_k^2} X_k X_k^\dagger.
\end{align}
To obtain Eq.~\eqref{master_LMG_tot}, we need to set the parameters such that
\begin{enumerate}
	\item $\Delta_0 = \Delta_1$ and hence $\delta_k^- = 0$,
	\item $\delta_{a_1}\gg\kappa_{a_1}$, $\delta_{b_1}\gg\kappa_{b_1}$, such that the effective dissipation terms for $a_1$ and $b_1$ can be neglected,
	\item $\alpha_{a_1} = \beta_{a_1} = \alpha_{a_2} = \beta_{a_2} = \alpha_{b_2} = \beta_{b_2} = \frac{1}{2}$, $\alpha_{b_1} = -\beta_{b_1} = \frac{1}{2\i}$, such that $X_k = S_x$ for $k=a_1,a_2,b_2$ and $X_{b_1} = S_y$,
	\item $-\lambda_{a_1}^2 \frac{\delta_{a_1}}{\kappa_{a_1}^2 + \delta_{a_1}^2} = \lambda_{b_1}^2 \frac{\delta_{b_1}}{\kappa_{b_1}^2 + \delta_{b_1}^2} = V$.
\end{enumerate}
The remaining parameters of Eq.~\eqref{master_LMG_tot} are then $h=\omega_0$, $\sqrt{(\gamma\kappa/2)}=-\lambda_{k}$, $\omega=\delta_k$, $\kappa = \kappa_k$ ($k=a_2$ or $k=b_2$), where one of the two modes $a_2$, $b_2$ can be removed from the dynamics by setting the corresponding Rabi frequencies to zero.

\section{Conclusion and outlook}
This work demonstrates that non-Markovianity can play a decisive role in the physics of dissipative phase transitions. By considering a non-Markovian extension of the maximally anisotropic Lipkin-Meshkov-Glick model with a transverse magnetic field, we have revealed the existence of a tricritical point that remains hidden in a purely Markovian description of the degrees of freedom of the spin. This finding underscores the rich interplay between quantum criticality and non-Markovianity, pointing to new directions for exploring dissipative phase transitions beyond the Markovian regime~\cite{Debecker2024C, Debecker2024S}. 

We have introduced the concept of Directional Spontaneous Symmetry Breaking (DSSB), which sheds light on multicritical phase diagrams, regardless of whether the dissipation is Markovian or non-Markovian. In particular, DSSB allows us to predict the number of phases that can be expected. In general, we believe that DSSB provides a theoretical framework that deepens our understanding of complex behavior in multicritical phase diagrams. An interesting future research direction would be to investigate the existence and properties of directional symmetry breaking of strong symmetries~\cite{Lieu2020S} and its implications on strong-to-weak symmetry breaking~\cite{Pablo2024S, Lessa2025S}.

In particular, we have shown that signatures of DSSB can be seen in our model from the emergence of steady-state spin squeezing along orthogonal directions. Elaborating on spin squeezing, we have then shown that non-Markovian dissipation can generate and enhance squeezing, thereby demonstrating that environments with memory can be seen as an active resource in tailoring quantum correlations and entanglement in driven-dissipative systems.

Finally, we have described a cavity-QED-based scheme to realize the non-Markovian LMG model~\eqref{master_LMG_tot} in the laboratory. By driving off-resonantly two ground-state and two excited-state levels of an ensemble of atoms with suitably chosen laser frequencies and coupling them off-resonantly to four cavity modes, one can adiabatically eliminate both the excited states and the faster cavity modes, leaving an effective Hamiltonian and dissipator that reproduce our LMG model. This setup should be within reach of current cavity-QED technology, offering a practical route for experimentally probing quantum phase transitions in driven-dissipative spin systems.

\acknowledgements
We thank Berislav Bu\v{c}a, Peter Kirton, Mattia Moroder, Kai Müller for useful comments on a previous version of the manuscript. B.D. thanks Eduardo Serrano-Ens\'astiga and Salvatore Araceli for fruitful discussions. J.M., T.B., and L.P. acknowledge the FWO and the F.R.S.-FNRS for their funding as part of the Excellence of Science program (EOS Project No. 40007526). T.B. also acknowledges financial support through IISN convention 4.4512.08. Computational resources were provided by the Consortium des Equipements de Calcul Intensif (CECI), funded by the Fonds de la Recherche Scientifique de Belgique (F.R.S.-FNRS) under Grant No. 2.5020.11.
\appendix 
\section{Exact computation of the squeezing parameter in the thermodynamic limit}
\label{Sec:AppTQ}
In this section, we show how to find exactly the spin squeezing parameter~\eqref{Ueda} in each phase in the thermodynamic limit. Before employing the third quantization to diagonalize the Liouvillian in the thermodynamic limit, we derive the effective Hamiltonian and jump operator in each phase, using a standard procedure in closed systems~\cite{Ma2009F, Dusuel2005C, Emary2003C} that has been extended to open systems~\cite{Buca2019D, Morrison2008C, Fan2023C}. In phase~(II), we already know the form of the effective Liouvillian in the thermodynamic limit~[see Eq.~\eqref{master_LMG_HP}]. 

\subsection{Effective Hamiltonian in phase (III)}
In the thermodynamic limit $N \rightarrow +\infty$, phase (III) is such that $\braket{a}/\sqrt{N} = \braket{S_x}/N= 0$, while $\braket{S_z}/N \neq 0$ and two branches emerge for $\braket{S_y}/N$. The general idea is to choose one of the two branches and to rotate the spin operators around the $x$ axis such that the $z$ axis is aligned with the mean spin direction. Therefore, we define implicitly $\mathbf{\tilde{S}} = (\tilde{S}_x, \tilde{S}_y, \tilde{S}_z)^T$ through 
\begin{equation}
    \begin{pmatrix}
        S_x \\
        S_y \\
        S_z
    \end{pmatrix} = 
    \begin{pmatrix}
        1 & 0 & 0 \\
        0  & \cos\theta & \sin \theta \\
        0 & -\sin \theta & \cos \theta
    \end{pmatrix}
    \begin{pmatrix}
       \tilde{S}_x \\
        \tilde{S}_y \\
        \tilde{S}_z
    \end{pmatrix}
    \label{Stilde_III}
\end{equation}
with $\cos \theta  \equiv m =  h/V$ and $\sin \theta = \sqrt{1-m^2}$, such that $\braket{\tilde{S}_z} = -1$. Next, we apply the HP transform to the transformed spin operators, i.e., 
\begin{equation}
    \begin{split}
        \tilde{S}_+ &= \sqrt{N} b^\dagger \sqrt{1 - \frac{b^\dagger b}{N}},\\
        \tilde{S}_- &= \sqrt{N} \sqrt{1 - \frac{b^\dagger b}{N}} b, \\
        \tilde{S}_z &= b^\dagger b - \frac{N}{2}.
    \end{split}
    \label{AHP}
\end{equation}
Writing the Hamiltonian~\eqref{H_LMG_tot} as a function of the rotated operators $\tilde{S}_i$ ($i= x, y,z$), and expanding the square root $\sqrt{1- b^\dagger b/N}  = 1 + O(1/N)$ gives the effective Hamiltonian 
\begin{equation}
    H_\mathrm{HP}^{\mathrm{(III)}} = \omega_a a^\dagger a + \omega_b b^\dagger b + G_\mathrm{eff} (b+b^\dagger)(a+a^\dagger) + V_\mathrm{eff} (b^{\dagger 2} + b^2),
    \label{HIII}
\end{equation}
where the bosonic modes' frequencies, their effective coupling and the squeezing strength of the mode $b$ are respectively given by 
\begin{equation}
\begin{split}
        &\omega _a = \omega,  \quad \omega_b = \frac32 V(1-m^2) + hm ,\\
        &G_\mathrm{eff} = \frac12\sqrt{\frac{\gamma \kappa}{2}},\quad 
        V_\mathrm{eff} = \frac{V}{4}(1+m^2).
\end{split}    
\end{equation}
To obtain the Hamiltonian~\eqref{HIII}, we discarded all constant terms. Note that it does not contain terms proportional to $\sqrt{N} (b+b^\dagger)$, which is expected as we precisely did a rotation of the spin operators to align the mean-field magnetization with the $z$ axis. Note also that when we do not apply any rotation ($\theta = 0 $ or $ m = 1$), we recover the Hamiltonian $H_\mathrm{HP}^{\mathrm{(II)}}$, as expected.
\subsection{Effective Hamiltonian and jump operator in phase (I)}
As for phase (III), we define new rotated spin operators $\tilde{S}_i$ ($i=x, y, z$) through  
\begin{equation}
    \begin{pmatrix}
        S_x\\
        S_y\\
        S_z
    \end{pmatrix} = 
    \begin{pmatrix}
        \cos \theta & 0 & \sin \theta \\
        0 & 1 & 0 \\
        -\sin \theta & 0 & \cos \theta 
    \end{pmatrix}
    \begin{pmatrix}
        \tilde{S}_x \\
        \tilde{S}_y \\
        \tilde{S}_z
    \end{pmatrix},
    \label{Stilde_I}
\end{equation}
with $\cos \theta \equiv m = - h /(V-q_2 \gamma/2)$ and $\sin \theta = \sqrt{1-m^2}$. As the pseudomode $a$ becomes macroscopically populated, we also define $\tilde{a}$ via 
\begin{equation}
     a= \tilde{a} + \sqrt{N} \alpha,
\end{equation}
with 
\begin{equation}
    \alpha = \frac12 \sqrt{\frac{\gamma}{2\kappa}}\sqrt{1-m^2} (q_2+iq_1).
\end{equation}
Now, we write the Liouvillian $\mathcal{L}_M$~[given by Eq.~\eqref{def_LM}] as a function of the rotated spin operators~\eqref{Stilde_I} and the shifted pseudomode $\tilde{a}$. We obtain the effective Liouvillian $\mathcal{L}_\mathrm{HP}^\mathrm{(I)}$ given by 
\begin{equation}
   \mathcal{L}_\mathrm{HP}^\mathrm{(I)}[\bigcdot] = -i[H_\mathrm{HP}^\mathrm{(I)}, \bigcdot] + \kappa (2 \tilde{a} \bigcdot \tilde{a}^\dagger - \{\tilde{a}^\dagger \tilde{a}, \bigcdot \}),
\end{equation}
up to $O(N^{-1/2})$ corrections. The effective Hamiltonian in phase (I) reads 
\begin{multline}
    H_\mathrm{HP}^\mathrm{(I)} =  \omega_a \tilde{a}^\dagger \tilde{a} + \omega_b b^\dagger b + G_\mathrm{eff}(b+b^\dagger)(\tilde{a} + \tilde{a}^\dagger) + V_\mathrm{eff}(b^{\dagger^2} + b^2),
\end{multline}
with 
\begin{align}
    &\omega_a = \omega, \quad \omega_b = \frac32(m^2-1)V + hm + \sqrt{\frac{\gamma \kappa}{2}} \sqrt{1-m^2} (\alpha + \alpha^*),\nonumber \\
    &G_\mathrm{eff} = \frac{m}{2}\sqrt{\frac{\gamma \kappa}{2}}, \quad V_\mathrm{eff} = \frac{V}{4}(1+m^2). 
\end{align}
Again, note that there is no term proportional to $\sqrt{N}(\tilde{a} + \tilde{a}^\dagger)$ nor $\sqrt{N}(b+b^\dagger)$ and setting $m=1$ gives $\alpha(m=1) = 0$ and $H_\mathrm{HP}^\mathrm{(I)} (m=1) =  H_\mathrm{HP}^\mathrm{(II)}$. 

\subsection{Applying the third quantization to compute squeezing parameters}
The third quantization method~\cite{Prosen2008T} (here applied to bosons~\cite{Prosen2010Q}) is a powerful method, akin to a Bogoliubov transformation but in the space of superoperators, allowing an efficient diagonalization of quadratic Liouvillians. Below, we explain how it can be applied in our case to the computation of the squeezing parameter.

In terms of the rotated operators $\vec{\tilde{S}}$, the spin squeezing parameter reads~\cite{Lee2014D, Ma2011Q}
\begin{equation}
    \xi^2 = \frac{2}{N}\left[\braket{\tilde{S}_{x}^2 + \tilde{S}_{y}^2} - \sqrt{\braket{\tilde{S}_{x}^2 - \tilde{S}_{y}^2}^2 + \braket{\{\tilde{S}_{x},\tilde{S}_{y}\}}^2}\right],
    \label{eq_appendix_xi2}
\end{equation}
with $\tilde{\vec S}$ given by Eq.~\eqref{Stilde_I} [Eq.~\eqref{Stilde_III}] in phase (I) [(III)] and  $\tilde{\vec S} = \vec S$ in phase (II). Here, we want to evaluate this expression exactly in the thermodynamic limit.

The effective Hamiltonian and jump operators evaluated in each phase leads to a quadratic Liouvillian whose Hamiltonian and jump operator adopt the form 
\begin{equation}
 \begin{split}
   H &=  \underline{c}^\dagger \cdot \boldsymbol{H} \underline{c}  + \underline{c} \cdot \boldsymbol{K} \underline{c} + \underline{c}^\dagger \cdot \boldsymbol{K}^*\underline{c}^\dagger , \\
   L &= \underline{l}\cdot \underline{c},
   \end{split}
\end{equation}
where we switched to standard notations in third quantization, i.e., matrices are denoted by bold symbols, $\underline{c} \equiv (b, \tilde{a})^T$, $\underline{c} \cdot \underline{d} = c_1d_1 + c_2d_2 $ defines a dot product between $\underline{c} = (c_1, c_2)^T$ and $\underline{d} = (d_1, d_2)^T$ and $\boldsymbol{K}^*$ is the conjugate matrix of $\boldsymbol{K}$. Note that $\boldsymbol{H}$ is Hermitian while $\boldsymbol{K}$ is symmetric. 

In this context, the steady state is Gaussian and is fully characterized by its second-order moments 
\begin{equation}
 Z_{ij} = \mathrm{Tr}[:d_i d_j: \rho_{\mathrm{ss}}], \quad \underline{d} = (\underline{c}, \underline{c}^\dagger),
\end{equation}
where $:d_id_j:$ denotes normal ordering such that e.g., $\braket{b^2} = Z_{11}$ and $\braket{b^\dagger b} = Z_{31}$. The correlation matrix $\boldsymbol{Z}$ is the $4 \times 4$ matrix  solution of the Lyapunov equation
\begin{equation}
    \boldsymbol{X}^T \boldsymbol{Z} + \boldsymbol{Z}\boldsymbol{X} = \boldsymbol{Y},
    \label{Lyapunov}
\end{equation}
where 
\begin{equation}
    \boldsymbol{X} \equiv \frac12 \begin{pmatrix}
       i \boldsymbol{H}^* + \boldsymbol{M} &-2 i \boldsymbol{K} \\
        2 i \boldsymbol{K}^* &  -i \boldsymbol{H} + \boldsymbol{M}^*
    \end{pmatrix}, \quad 
    \boldsymbol{Y} \equiv \frac12
    \begin{pmatrix}
       -2 i \boldsymbol{K}^*  & \boldsymbol{0}_{2\times 2}\\
       \boldsymbol{0}_{2\times 2}&  2 i \boldsymbol{K}
    \end{pmatrix},   
\end{equation}
and 
\begin{equation}
   \boldsymbol{M} = \underline{l}^*\underline{l}^T = \begin{pmatrix}
             0 & 0 \\
             0 & \kappa
   \end{pmatrix}.
\end{equation}
Evaluating the squeezing parameter~\eqref{eq_appendix_xi2} is now a straightforward procedure. In each phase, we identify  the matrices $\boldsymbol{H}, \boldsymbol{K}$ and $\boldsymbol{K^*}$. Then we obtain the correlation matrix $\boldsymbol{Z}$ by solving the matrix equation~\eqref{Lyapunov}. Finally, we plug the fluctuations of the Holstein-Primakoff boson in Eq.~\eqref{eq_appendix_xi2}.

To be abundantly explicit, we close this section by giving the matrices $\boldsymbol{H}$ and $\boldsymbol{K}$ in phase (II). They read 
\begin{equation}
   \boldsymbol{H} = \begin{pmatrix}
       h & \frac12 \sqrt{\gamma\kappa/2} \\
       \frac12\sqrt{\gamma\kappa/2} & \omega 
   \end{pmatrix}, \quad \boldsymbol{K} = \begin{pmatrix}
       V/2 & \frac14 \sqrt{\gamma\kappa/2} \\
       \frac14\sqrt{\gamma\kappa/2} & 0 
   \end{pmatrix}.
\end{equation}
\bibliographystyle{apsrev4-2}
\bibliography{bib}

\begin{thebibliography}{81}%
\makeatletter
\providecommand \@ifxundefined [1]{%
 \@ifx{#1\undefined}
}%
\providecommand \@ifnum [1]{%
 \ifnum #1\expandafter \@firstoftwo
 \else \expandafter \@secondoftwo
 \fi
}%
\providecommand \@ifx [1]{%
 \ifx #1\expandafter \@firstoftwo
 \else \expandafter \@secondoftwo
 \fi
}%
\providecommand \natexlab [1]{#1}%
\providecommand \enquote  [1]{``#1''}%
\providecommand \bibnamefont  [1]{#1}%
\providecommand \bibfnamefont [1]{#1}%
\providecommand \citenamefont [1]{#1}%
\providecommand \href@noop [0]{\@secondoftwo}%
\providecommand \href [0]{\begingroup \@sanitize@url \@href}%
\providecommand \@href[1]{\@@startlink{#1}\@@href}%
\providecommand \@@href[1]{\endgroup#1\@@endlink}%
\providecommand \@sanitize@url [0]{\catcode `\\12\catcode `\$12\catcode `\&12\catcode `\#12\catcode `\^12\catcode `\_12\catcode `\%12\relax}%
\providecommand \@@startlink[1]{}%
\providecommand \@@endlink[0]{}%
\providecommand \url  [0]{\begingroup\@sanitize@url \@url }%
\providecommand \@url [1]{\endgroup\@href {#1}{\urlprefix }}%
\providecommand \urlprefix  [0]{URL }%
\providecommand \Eprint [0]{\href }%
\providecommand \doibase [0]{https://doi.org/}%
\providecommand \selectlanguage [0]{\@gobble}%
\providecommand \bibinfo  [0]{\@secondoftwo}%
\providecommand \bibfield  [0]{\@secondoftwo}%
\providecommand \translation [1]{[#1]}%
\providecommand \BibitemOpen [0]{}%
\providecommand \bibitemStop [0]{}%
\providecommand \bibitemNoStop [0]{.\EOS\space}%
\providecommand \EOS [0]{\spacefactor3000\relax}%
\providecommand \BibitemShut  [1]{\csname bibitem#1\endcsname}%
\let\auto@bib@innerbib\@empty
%</preamble>
\bibitem [{\citenamefont {Roberts}\ and\ \citenamefont {Clerk}(2020)}]{Roberts2020D}%
  \BibitemOpen
  \bibfield  {author} {\bibinfo {author} {\bibfnamefont {D.}~\bibnamefont {Roberts}}\ and\ \bibinfo {author} {\bibfnamefont {A.~A.}\ \bibnamefont {Clerk}},\ }\href {https://doi.org/10.1103/PhysRevX.10.021022} {\bibfield  {journal} {\bibinfo  {journal} {Phys. Rev. X}\ }\textbf {\bibinfo {volume} {10}},\ \bibinfo {pages} {021022} (\bibinfo {year} {2020})}\BibitemShut {NoStop}%
\bibitem [{\citenamefont {Zapletal}\ \emph {et~al.}(2022)\citenamefont {Zapletal}, \citenamefont {Nunnenkamp},\ and\ \citenamefont {Brunelli}}]{Zapletal2022S}%
  \BibitemOpen
  \bibfield  {author} {\bibinfo {author} {\bibfnamefont {P.}~\bibnamefont {Zapletal}}, \bibinfo {author} {\bibfnamefont {A.}~\bibnamefont {Nunnenkamp}},\ and\ \bibinfo {author} {\bibfnamefont {M.}~\bibnamefont {Brunelli}},\ }\href {https://doi.org/10.1103/PRXQuantum.3.010301} {\bibfield  {journal} {\bibinfo  {journal} {PRX Quantum}\ }\textbf {\bibinfo {volume} {3}},\ \bibinfo {pages} {010301} (\bibinfo {year} {2022})}\BibitemShut {NoStop}%
\bibitem [{\citenamefont {Toner}\ and\ \citenamefont {Tu}(1998)}]{Toner1998Flocks}%
  \BibitemOpen
  \bibfield  {author} {\bibinfo {author} {\bibfnamefont {J.}~\bibnamefont {Toner}}\ and\ \bibinfo {author} {\bibfnamefont {Y.}~\bibnamefont {Tu}},\ }\href {https://doi.org/10.1103/PhysRevE.58.4828} {\bibfield  {journal} {\bibinfo  {journal} {Phys. Rev. E}\ }\textbf {\bibinfo {volume} {58}},\ \bibinfo {pages} {4828} (\bibinfo {year} {1998})}\BibitemShut {NoStop}%
\bibitem [{\citenamefont {Jin}\ \emph {et~al.}(2016)\citenamefont {Jin}, \citenamefont {Biella}, \citenamefont {Viyuela}, \citenamefont {Mazza}, \citenamefont {Keeling}, \citenamefont {Fazio},\ and\ \citenamefont {Rossini}}]{Jin2016Cluster}%
  \BibitemOpen
  \bibfield  {author} {\bibinfo {author} {\bibfnamefont {J.}~\bibnamefont {Jin}}, \bibinfo {author} {\bibfnamefont {A.}~\bibnamefont {Biella}}, \bibinfo {author} {\bibfnamefont {O.}~\bibnamefont {Viyuela}}, \bibinfo {author} {\bibfnamefont {L.}~\bibnamefont {Mazza}}, \bibinfo {author} {\bibfnamefont {J.}~\bibnamefont {Keeling}}, \bibinfo {author} {\bibfnamefont {R.}~\bibnamefont {Fazio}},\ and\ \bibinfo {author} {\bibfnamefont {D.}~\bibnamefont {Rossini}},\ }\href {https://doi.org/10.1103/PhysRevX.6.031011} {\bibfield  {journal} {\bibinfo  {journal} {Phys. Rev. X}\ }\textbf {\bibinfo {volume} {6}},\ \bibinfo {pages} {031011} (\bibinfo {year} {2016})}\BibitemShut {NoStop}%
\bibitem [{\citenamefont {Lee}\ \emph {et~al.}(2013)\citenamefont {Lee}, \citenamefont {Gopalakrishnan},\ and\ \citenamefont {Lukin}}]{Lee2013Unconventional}%
  \BibitemOpen
  \bibfield  {author} {\bibinfo {author} {\bibfnamefont {T.~E.}\ \bibnamefont {Lee}}, \bibinfo {author} {\bibfnamefont {S.}~\bibnamefont {Gopalakrishnan}},\ and\ \bibinfo {author} {\bibfnamefont {M.~D.}\ \bibnamefont {Lukin}},\ }\href {https://doi.org/10.1103/PhysRevLett.110.257204} {\bibfield  {journal} {\bibinfo  {journal} {Phys. Rev. Lett.}\ }\textbf {\bibinfo {volume} {110}},\ \bibinfo {pages} {257204} (\bibinfo {year} {2013})}\BibitemShut {NoStop}%
\bibitem [{\citenamefont {Diehl}\ \emph {et~al.}(2008)\citenamefont {Diehl}, \citenamefont {Micheli}, \citenamefont {Kantian}, \citenamefont {Kraus}, \citenamefont {B{\"u}chler},\ and\ \citenamefont {Zoller}}]{Diehl2008Q}%
  \BibitemOpen
  \bibfield  {author} {\bibinfo {author} {\bibfnamefont {S.}~\bibnamefont {Diehl}}, \bibinfo {author} {\bibfnamefont {A.}~\bibnamefont {Micheli}}, \bibinfo {author} {\bibfnamefont {A.}~\bibnamefont {Kantian}}, \bibinfo {author} {\bibfnamefont {B.}~\bibnamefont {Kraus}}, \bibinfo {author} {\bibfnamefont {H.~P.}\ \bibnamefont {B{\"u}chler}},\ and\ \bibinfo {author} {\bibfnamefont {P.}~\bibnamefont {Zoller}},\ }\href {https://doi.org/10.1038/nphys1073} {\bibfield  {journal} {\bibinfo  {journal} {Nature Physics}\ }\textbf {\bibinfo {volume} {4}},\ \bibinfo {pages} {878} (\bibinfo {year} {2008})}\BibitemShut {NoStop}%
\bibitem [{\citenamefont {Verstraete}\ \emph {et~al.}(2009)\citenamefont {Verstraete}, \citenamefont {Wolf},\ and\ \citenamefont {Ignacio~Cirac}}]{Verstraete2009Q}%
  \BibitemOpen
  \bibfield  {author} {\bibinfo {author} {\bibfnamefont {F.}~\bibnamefont {Verstraete}}, \bibinfo {author} {\bibfnamefont {M.~M.}\ \bibnamefont {Wolf}},\ and\ \bibinfo {author} {\bibfnamefont {J.}~\bibnamefont {Ignacio~Cirac}},\ }\href {https://doi.org/10.1038/nphys1342} {\bibfield  {journal} {\bibinfo  {journal} {Nature Physics}\ }\textbf {\bibinfo {volume} {5}},\ \bibinfo {pages} {633} (\bibinfo {year} {2009})}\BibitemShut {NoStop}%
\bibitem [{\citenamefont {Koppenh\"ofer}\ \emph {et~al.}(2022)\citenamefont {Koppenh\"ofer}, \citenamefont {Groszkowski}, \citenamefont {Lau},\ and\ \citenamefont {Clerk}}]{Koppenhofer2022D}%
  \BibitemOpen
  \bibfield  {author} {\bibinfo {author} {\bibfnamefont {M.}~\bibnamefont {Koppenh\"ofer}}, \bibinfo {author} {\bibfnamefont {P.}~\bibnamefont {Groszkowski}}, \bibinfo {author} {\bibfnamefont {H.-K.}\ \bibnamefont {Lau}},\ and\ \bibinfo {author} {\bibfnamefont {A.}~\bibnamefont {Clerk}},\ }\href {https://doi.org/10.1103/PRXQuantum.3.030330} {\bibfield  {journal} {\bibinfo  {journal} {PRX Quantum}\ }\textbf {\bibinfo {volume} {3}},\ \bibinfo {pages} {030330} (\bibinfo {year} {2022})}\BibitemShut {NoStop}%
\bibitem [{\citenamefont {Yang}\ \emph {et~al.}(2024)\citenamefont {Yang}, \citenamefont {Long}, \citenamefont {Liu}, \citenamefont {Tang}, \citenamefont {Zhai}, \citenamefont {Nie}, \citenamefont {Xin}, \citenamefont {Li},\ and\ \citenamefont {Lu}}]{Yang2024C}%
  \BibitemOpen
  \bibfield  {author} {\bibinfo {author} {\bibfnamefont {X.}~\bibnamefont {Yang}}, \bibinfo {author} {\bibfnamefont {X.}~\bibnamefont {Long}}, \bibinfo {author} {\bibfnamefont {R.}~\bibnamefont {Liu}}, \bibinfo {author} {\bibfnamefont {K.}~\bibnamefont {Tang}}, \bibinfo {author} {\bibfnamefont {Y.}~\bibnamefont {Zhai}}, \bibinfo {author} {\bibfnamefont {X.}~\bibnamefont {Nie}}, \bibinfo {author} {\bibfnamefont {T.}~\bibnamefont {Xin}}, \bibinfo {author} {\bibfnamefont {J.}~\bibnamefont {Li}},\ and\ \bibinfo {author} {\bibfnamefont {D.}~\bibnamefont {Lu}},\ }\href {https://doi.org/10.1038/s42005-024-01758-8} {\bibfield  {journal} {\bibinfo  {journal} {Communications Physics}\ }\textbf {\bibinfo {volume} {7}},\ \bibinfo {pages} {282} (\bibinfo {year} {2024})}\BibitemShut {NoStop}%
\bibitem [{\citenamefont {de~Vega}\ and\ \citenamefont {Alonso}(2017)}]{deVega2017Dynamics}%
  \BibitemOpen
  \bibfield  {author} {\bibinfo {author} {\bibfnamefont {I.}~\bibnamefont {de~Vega}}\ and\ \bibinfo {author} {\bibfnamefont {D.}~\bibnamefont {Alonso}},\ }\href {https://doi.org/10.1103/RevModPhys.89.015001} {\bibfield  {journal} {\bibinfo  {journal} {Rev. Mod. Phys.}\ }\textbf {\bibinfo {volume} {89}},\ \bibinfo {pages} {015001} (\bibinfo {year} {2017})}\BibitemShut {NoStop}%
\bibitem [{\citenamefont {Gr{\"o}blacher}\ \emph {et~al.}(2015)\citenamefont {Gr{\"o}blacher}, \citenamefont {Trubarov}, \citenamefont {Prigge}, \citenamefont {Cole}, \citenamefont {Aspelmeyer},\ and\ \citenamefont {Eisert}}]{Groblacher2015}%
  \BibitemOpen
  \bibfield  {author} {\bibinfo {author} {\bibfnamefont {S.}~\bibnamefont {Gr{\"o}blacher}}, \bibinfo {author} {\bibfnamefont {A.}~\bibnamefont {Trubarov}}, \bibinfo {author} {\bibfnamefont {N.}~\bibnamefont {Prigge}}, \bibinfo {author} {\bibfnamefont {G.~D.}\ \bibnamefont {Cole}}, \bibinfo {author} {\bibfnamefont {M.}~\bibnamefont {Aspelmeyer}},\ and\ \bibinfo {author} {\bibfnamefont {J.}~\bibnamefont {Eisert}},\ }\href {https://doi.org/10.1038/ncomms8606} {\bibfield  {journal} {\bibinfo  {journal} {Nature Communications}\ }\textbf {\bibinfo {volume} {6}},\ \bibinfo {pages} {7606} (\bibinfo {year} {2015})}\BibitemShut {NoStop}%
\bibitem [{\citenamefont {Poto{\v{c}}nik}\ \emph {et~al.}(2018)\citenamefont {Poto{\v{c}}nik}, \citenamefont {Bargerbos}, \citenamefont {Schr{\"o}der}, \citenamefont {Khan}, \citenamefont {Collodo}, \citenamefont {Gasparinetti}, \citenamefont {Salath{\'e}}, \citenamefont {Creatore}, \citenamefont {Eichler}, \citenamefont {T{\"u}reci}, \citenamefont {Chin},\ and\ \citenamefont {Wallraff}}]{Potocnik2018}%
  \BibitemOpen
  \bibfield  {author} {\bibinfo {author} {\bibfnamefont {A.}~\bibnamefont {Poto{\v{c}}nik}}, \bibinfo {author} {\bibfnamefont {A.}~\bibnamefont {Bargerbos}}, \bibinfo {author} {\bibfnamefont {F.~A. Y.~N.}\ \bibnamefont {Schr{\"o}der}}, \bibinfo {author} {\bibfnamefont {S.~A.}\ \bibnamefont {Khan}}, \bibinfo {author} {\bibfnamefont {M.~C.}\ \bibnamefont {Collodo}}, \bibinfo {author} {\bibfnamefont {S.}~\bibnamefont {Gasparinetti}}, \bibinfo {author} {\bibfnamefont {Y.}~\bibnamefont {Salath{\'e}}}, \bibinfo {author} {\bibfnamefont {C.}~\bibnamefont {Creatore}}, \bibinfo {author} {\bibfnamefont {C.}~\bibnamefont {Eichler}}, \bibinfo {author} {\bibfnamefont {H.~E.}\ \bibnamefont {T{\"u}reci}}, \bibinfo {author} {\bibfnamefont {A.~W.}\ \bibnamefont {Chin}},\ and\ \bibinfo {author} {\bibfnamefont {A.}~\bibnamefont {Wallraff}},\ }\href {https://doi.org/10.1038/s41467-018-03312-x} {\bibfield  {journal} {\bibinfo  {journal} {Nature Communications}\ }\textbf {\bibinfo {volume} {9}},\ \bibinfo {pages} {904} (\bibinfo
  {year} {2018})}\BibitemShut {NoStop}%
\bibitem [{\citenamefont {Madsen}\ \emph {et~al.}(2011)\citenamefont {Madsen}, \citenamefont {Ates}, \citenamefont {Lund-Hansen}, \citenamefont {L\"offler}, \citenamefont {Reitzenstein}, \citenamefont {Forchel},\ and\ \citenamefont {Lodahl}}]{Madsen2011}%
  \BibitemOpen
  \bibfield  {author} {\bibinfo {author} {\bibfnamefont {K.~H.}\ \bibnamefont {Madsen}}, \bibinfo {author} {\bibfnamefont {S.}~\bibnamefont {Ates}}, \bibinfo {author} {\bibfnamefont {T.}~\bibnamefont {Lund-Hansen}}, \bibinfo {author} {\bibfnamefont {A.}~\bibnamefont {L\"offler}}, \bibinfo {author} {\bibfnamefont {S.}~\bibnamefont {Reitzenstein}}, \bibinfo {author} {\bibfnamefont {A.}~\bibnamefont {Forchel}},\ and\ \bibinfo {author} {\bibfnamefont {P.}~\bibnamefont {Lodahl}},\ }\href {https://doi.org/10.1103/PhysRevLett.106.233601} {\bibfield  {journal} {\bibinfo  {journal} {Phys. Rev. Lett.}\ }\textbf {\bibinfo {volume} {106}},\ \bibinfo {pages} {233601} (\bibinfo {year} {2011})}\BibitemShut {NoStop}%
\bibitem [{\citenamefont {Vicencio}\ \emph {et~al.}(2025)\citenamefont {Vicencio}, \citenamefont {Carcamo-Macaya},\ and\ \citenamefont {Solano}}]{Vicencio2025}%
  \BibitemOpen
  \bibfield  {author} {\bibinfo {author} {\bibfnamefont {R.~A.}\ \bibnamefont {Vicencio}}, \bibinfo {author} {\bibfnamefont {F.~G.~L.}\ \bibnamefont {Carcamo-Macaya}},\ and\ \bibinfo {author} {\bibfnamefont {P.}~\bibnamefont {Solano}},\ }\href {https://arxiv.org/abs/2501.09261} {\bibinfo {title} {Observation of non-markovian radiative phenomena in structured photonic lattices}} (\bibinfo {year} {2025}),\ \Eprint {https://arxiv.org/abs/2501.09261} {arXiv:2501.09261 [quant-ph]} \BibitemShut {NoStop}%
\bibitem [{\citenamefont {Palacino}\ and\ \citenamefont {Keeling}(2021)}]{Palacino2020}%
  \BibitemOpen
  \bibfield  {author} {\bibinfo {author} {\bibfnamefont {R.}~\bibnamefont {Palacino}}\ and\ \bibinfo {author} {\bibfnamefont {J.}~\bibnamefont {Keeling}},\ }\href {https://doi.org/10.1103/PhysRevResearch.3.L032016} {\bibfield  {journal} {\bibinfo  {journal} {Phys. Rev. Research}\ }\textbf {\bibinfo {volume} {3}},\ \bibinfo {pages} {L032016} (\bibinfo {year} {2021})}\BibitemShut {NoStop}%
\bibitem [{\citenamefont {Damanet}\ \emph {et~al.}(2019)\citenamefont {Damanet}, \citenamefont {Daley},\ and\ \citenamefont {Keeling}}]{Damanet2019}%
  \BibitemOpen
  \bibfield  {author} {\bibinfo {author} {\bibfnamefont {F.}~\bibnamefont {Damanet}}, \bibinfo {author} {\bibfnamefont {A.~J.}\ \bibnamefont {Daley}},\ and\ \bibinfo {author} {\bibfnamefont {J.}~\bibnamefont {Keeling}},\ }\href {https://doi.org/10.1103/PhysRevA.99.033845} {\bibfield  {journal} {\bibinfo  {journal} {Phys. Rev. A}\ }\textbf {\bibinfo {volume} {99}},\ \bibinfo {pages} {033845} (\bibinfo {year} {2019})}\BibitemShut {NoStop}%
\bibitem [{\citenamefont {Link}\ \emph {et~al.}(2022)\citenamefont {Link}, \citenamefont {M\"uller}, \citenamefont {Lena}, \citenamefont {Luoma}, \citenamefont {Damanet}, \citenamefont {Strunz},\ and\ \citenamefont {Daley}}]{Link2022}%
  \BibitemOpen
  \bibfield  {author} {\bibinfo {author} {\bibfnamefont {V.}~\bibnamefont {Link}}, \bibinfo {author} {\bibfnamefont {K.}~\bibnamefont {M\"uller}}, \bibinfo {author} {\bibfnamefont {R.~G.}\ \bibnamefont {Lena}}, \bibinfo {author} {\bibfnamefont {K.}~\bibnamefont {Luoma}}, \bibinfo {author} {\bibfnamefont {F.}~\bibnamefont {Damanet}}, \bibinfo {author} {\bibfnamefont {W.~T.}\ \bibnamefont {Strunz}},\ and\ \bibinfo {author} {\bibfnamefont {A.~J.}\ \bibnamefont {Daley}},\ }\href {https://doi.org/10.1103/PRXQuantum.3.020348} {\bibfield  {journal} {\bibinfo  {journal} {PRX Quantum}\ }\textbf {\bibinfo {volume} {3}},\ \bibinfo {pages} {020348} (\bibinfo {year} {2022})}\BibitemShut {NoStop}%
\bibitem [{\citenamefont {Ask}\ and\ \citenamefont {Johansson}(2022)}]{Ask2022}%
  \BibitemOpen
  \bibfield  {author} {\bibinfo {author} {\bibfnamefont {A.}~\bibnamefont {Ask}}\ and\ \bibinfo {author} {\bibfnamefont {G.}~\bibnamefont {Johansson}},\ }\href {https://doi.org/10.1103/PhysRevLett.128.083603} {\bibfield  {journal} {\bibinfo  {journal} {Phys. Rev. Lett.}\ }\textbf {\bibinfo {volume} {128}},\ \bibinfo {pages} {083603} (\bibinfo {year} {2022})}\BibitemShut {NoStop}%
\bibitem [{\citenamefont {Huelga}\ \emph {et~al.}(2012)\citenamefont {Huelga}, \citenamefont {Rivas},\ and\ \citenamefont {Plenio}}]{Huelga2012Non}%
  \BibitemOpen
  \bibfield  {author} {\bibinfo {author} {\bibfnamefont {S.~F.}\ \bibnamefont {Huelga}}, \bibinfo {author} {\bibfnamefont {A.}~\bibnamefont {Rivas}},\ and\ \bibinfo {author} {\bibfnamefont {M.~B.}\ \bibnamefont {Plenio}},\ }\href {https://doi.org/10.1103/PhysRevLett.108.160402} {\bibfield  {journal} {\bibinfo  {journal} {Phys. Rev. Lett.}\ }\textbf {\bibinfo {volume} {108}},\ \bibinfo {pages} {160402} (\bibinfo {year} {2012})}\BibitemShut {NoStop}%
\bibitem [{\citenamefont {Maier}\ \emph {et~al.}(2019)\citenamefont {Maier}, \citenamefont {Brydges}, \citenamefont {Jurcevic}, \citenamefont {Trautmann}, \citenamefont {Hempel}, \citenamefont {Lanyon}, \citenamefont {Hauke}, \citenamefont {Blatt},\ and\ \citenamefont {Roos}}]{Maier2019}%
  \BibitemOpen
  \bibfield  {author} {\bibinfo {author} {\bibfnamefont {C.}~\bibnamefont {Maier}}, \bibinfo {author} {\bibfnamefont {T.}~\bibnamefont {Brydges}}, \bibinfo {author} {\bibfnamefont {P.}~\bibnamefont {Jurcevic}}, \bibinfo {author} {\bibfnamefont {N.}~\bibnamefont {Trautmann}}, \bibinfo {author} {\bibfnamefont {C.}~\bibnamefont {Hempel}}, \bibinfo {author} {\bibfnamefont {B.~P.}\ \bibnamefont {Lanyon}}, \bibinfo {author} {\bibfnamefont {P.}~\bibnamefont {Hauke}}, \bibinfo {author} {\bibfnamefont {R.}~\bibnamefont {Blatt}},\ and\ \bibinfo {author} {\bibfnamefont {C.~F.}\ \bibnamefont {Roos}},\ }\href {https://doi.org/10.1103/PhysRevLett.122.050501} {\bibfield  {journal} {\bibinfo  {journal} {Phys. Rev. Lett.}\ }\textbf {\bibinfo {volume} {122}},\ \bibinfo {pages} {050501} (\bibinfo {year} {2019})}\BibitemShut {NoStop}%
\bibitem [{\citenamefont {Kuo}\ \emph {et~al.}(2024)\citenamefont {Kuo}, \citenamefont {Yang}, \citenamefont {Lambert}, \citenamefont {Lin}, \citenamefont {Huang}, \citenamefont {Nori},\ and\ \citenamefont {Chen}}]{Kuo2024N}%
  \BibitemOpen
  \bibfield  {author} {\bibinfo {author} {\bibfnamefont {P.-C.}\ \bibnamefont {Kuo}}, \bibinfo {author} {\bibfnamefont {S.-L.}\ \bibnamefont {Yang}}, \bibinfo {author} {\bibfnamefont {N.}~\bibnamefont {Lambert}}, \bibinfo {author} {\bibfnamefont {J.-D.}\ \bibnamefont {Lin}}, \bibinfo {author} {\bibfnamefont {Y.-T.}\ \bibnamefont {Huang}}, \bibinfo {author} {\bibfnamefont {F.}~\bibnamefont {Nori}},\ and\ \bibinfo {author} {\bibfnamefont {Y.-N.}\ \bibnamefont {Chen}},\ }\href {https://arxiv.org/abs/2403.14455} {\bibinfo {title} {Non-markovian skin effect}} (\bibinfo {year} {2024}),\ \Eprint {https://arxiv.org/abs/2403.14455} {arXiv:2403.14455 [quant-ph]} \BibitemShut {NoStop}%
\bibitem [{\citenamefont {Debecker}\ \emph {et~al.}(2024{\natexlab{a}})\citenamefont {Debecker}, \citenamefont {Martin},\ and\ \citenamefont {Damanet}}]{Debecker2024S}%
  \BibitemOpen
  \bibfield  {author} {\bibinfo {author} {\bibfnamefont {B.}~\bibnamefont {Debecker}}, \bibinfo {author} {\bibfnamefont {J.}~\bibnamefont {Martin}},\ and\ \bibinfo {author} {\bibfnamefont {F.}~\bibnamefont {Damanet}},\ }\href {https://doi.org/10.1103/PhysRevA.110.042201} {\bibfield  {journal} {\bibinfo  {journal} {Phys. Rev. A}\ }\textbf {\bibinfo {volume} {110}},\ \bibinfo {pages} {042201} (\bibinfo {year} {2024}{\natexlab{a}})}\BibitemShut {NoStop}%
\bibitem [{\citenamefont {Debecker}\ \emph {et~al.}(2024{\natexlab{b}})\citenamefont {Debecker}, \citenamefont {Martin},\ and\ \citenamefont {Damanet}}]{Debecker2024C}%
  \BibitemOpen
  \bibfield  {author} {\bibinfo {author} {\bibfnamefont {B.}~\bibnamefont {Debecker}}, \bibinfo {author} {\bibfnamefont {J.}~\bibnamefont {Martin}},\ and\ \bibinfo {author} {\bibfnamefont {F.}~\bibnamefont {Damanet}},\ }\href {https://doi.org/10.1103/PhysRevLett.133.140403} {\bibfield  {journal} {\bibinfo  {journal} {Phys. Rev. Lett.}\ }\textbf {\bibinfo {volume} {133}},\ \bibinfo {pages} {140403} (\bibinfo {year} {2024}{\natexlab{b}})}\BibitemShut {NoStop}%
\bibitem [{\citenamefont {Soriente}\ \emph {et~al.}(2018)\citenamefont {Soriente}, \citenamefont {Donner}, \citenamefont {Chitra},\ and\ \citenamefont {Zilberberg}}]{Soriente2018D}%
  \BibitemOpen
  \bibfield  {author} {\bibinfo {author} {\bibfnamefont {M.}~\bibnamefont {Soriente}}, \bibinfo {author} {\bibfnamefont {T.}~\bibnamefont {Donner}}, \bibinfo {author} {\bibfnamefont {R.}~\bibnamefont {Chitra}},\ and\ \bibinfo {author} {\bibfnamefont {O.}~\bibnamefont {Zilberberg}},\ }\href {https://doi.org/10.1103/PhysRevLett.120.183603} {\bibfield  {journal} {\bibinfo  {journal} {Phys. Rev. Lett.}\ }\textbf {\bibinfo {volume} {120}},\ \bibinfo {pages} {183603} (\bibinfo {year} {2018})}\BibitemShut {NoStop}%
\bibitem [{\citenamefont {Lipkin}\ \emph {et~al.}(1965)\citenamefont {Lipkin}, \citenamefont {Meshkov},\ and\ \citenamefont {Glick}}]{Lipkin1965V}%
  \BibitemOpen
  \bibfield  {author} {\bibinfo {author} {\bibfnamefont {H.}~\bibnamefont {Lipkin}}, \bibinfo {author} {\bibfnamefont {N.}~\bibnamefont {Meshkov}},\ and\ \bibinfo {author} {\bibfnamefont {A.}~\bibnamefont {Glick}},\ }\href {https://doi.org/https://doi.org/10.1016/0029-5582(65)90862-X} {\bibfield  {journal} {\bibinfo  {journal} {Nuclear Physics}\ }\textbf {\bibinfo {volume} {62}},\ \bibinfo {pages} {188} (\bibinfo {year} {1965})}\BibitemShut {NoStop}%
\bibitem [{\citenamefont {Vidal}\ \emph {et~al.}(2004)\citenamefont {Vidal}, \citenamefont {Palacios},\ and\ \citenamefont {Aslangul}}]{Vidal2004E}%
  \BibitemOpen
  \bibfield  {author} {\bibinfo {author} {\bibfnamefont {J.}~\bibnamefont {Vidal}}, \bibinfo {author} {\bibfnamefont {G.}~\bibnamefont {Palacios}},\ and\ \bibinfo {author} {\bibfnamefont {C.}~\bibnamefont {Aslangul}},\ }\href {https://doi.org/10.1103/PhysRevA.70.062304} {\bibfield  {journal} {\bibinfo  {journal} {Phys. Rev. A}\ }\textbf {\bibinfo {volume} {70}},\ \bibinfo {pages} {062304} (\bibinfo {year} {2004})}\BibitemShut {NoStop}%
\bibitem [{\citenamefont {Wilms}\ \emph {et~al.}(2012)\citenamefont {Wilms}, \citenamefont {Vidal}, \citenamefont {Verstraete},\ and\ \citenamefont {Dusuel}}]{Wilms2012J}%
  \BibitemOpen
  \bibfield  {author} {\bibinfo {author} {\bibfnamefont {J.}~\bibnamefont {Wilms}}, \bibinfo {author} {\bibfnamefont {J.}~\bibnamefont {Vidal}}, \bibinfo {author} {\bibfnamefont {F.}~\bibnamefont {Verstraete}},\ and\ \bibinfo {author} {\bibfnamefont {S.}~\bibnamefont {Dusuel}},\ }\href {https://doi.org/10.1088/1742-5468/2012/01/P01023} {\bibfield  {journal} {\bibinfo  {journal} {Journal of Statistical Mechanics: Theory and Experiment}\ }\textbf {\bibinfo {volume} {2012}},\ \bibinfo {pages} {P01023} (\bibinfo {year} {2012})}\BibitemShut {NoStop}%
\bibitem [{\citenamefont {Ribeiro}\ \emph {et~al.}(2007)\citenamefont {Ribeiro}, \citenamefont {Vidal},\ and\ \citenamefont {Mosseri}}]{Pedro2007T}%
  \BibitemOpen
  \bibfield  {author} {\bibinfo {author} {\bibfnamefont {P.}~\bibnamefont {Ribeiro}}, \bibinfo {author} {\bibfnamefont {J.}~\bibnamefont {Vidal}},\ and\ \bibinfo {author} {\bibfnamefont {R.}~\bibnamefont {Mosseri}},\ }\href {https://doi.org/10.1103/PhysRevLett.99.050402} {\bibfield  {journal} {\bibinfo  {journal} {Phys. Rev. Lett.}\ }\textbf {\bibinfo {volume} {99}},\ \bibinfo {pages} {050402} (\bibinfo {year} {2007})}\BibitemShut {NoStop}%
\bibitem [{\citenamefont {Or\'us}\ \emph {et~al.}(2008)\citenamefont {Or\'us}, \citenamefont {Dusuel},\ and\ \citenamefont {Vidal}}]{Orus2008E}%
  \BibitemOpen
  \bibfield  {author} {\bibinfo {author} {\bibfnamefont {R.}~\bibnamefont {Or\'us}}, \bibinfo {author} {\bibfnamefont {S.}~\bibnamefont {Dusuel}},\ and\ \bibinfo {author} {\bibfnamefont {J.}~\bibnamefont {Vidal}},\ }\href {https://doi.org/10.1103/PhysRevLett.101.025701} {\bibfield  {journal} {\bibinfo  {journal} {Phys. Rev. Lett.}\ }\textbf {\bibinfo {volume} {101}},\ \bibinfo {pages} {025701} (\bibinfo {year} {2008})}\BibitemShut {NoStop}%
\bibitem [{\citenamefont {Ptaszy\ifmmode~\acute{n}\else \'{n}\fi{}ski}\ and\ \citenamefont {Esposito}(2024)}]{Ptaszy2024D}%
  \BibitemOpen
  \bibfield  {author} {\bibinfo {author} {\bibfnamefont {K.}~\bibnamefont {Ptaszy\ifmmode~\acute{n}\else \'{n}\fi{}ski}}\ and\ \bibinfo {author} {\bibfnamefont {M.}~\bibnamefont {Esposito}},\ }\href {https://doi.org/10.1103/PhysRevE.110.044134} {\bibfield  {journal} {\bibinfo  {journal} {Phys. Rev. E}\ }\textbf {\bibinfo {volume} {110}},\ \bibinfo {pages} {044134} (\bibinfo {year} {2024})}\BibitemShut {NoStop}%
\bibitem [{\citenamefont {Breuer}\ and\ \citenamefont {Petruccione}(2006)}]{Bre06}%
  \BibitemOpen
  \bibfield  {author} {\bibinfo {author} {\bibfnamefont {H.-P.}\ \bibnamefont {Breuer}}\ and\ \bibinfo {author} {\bibfnamefont {F.}~\bibnamefont {Petruccione}},\ }\href@noop {} {\emph {\bibinfo {title} {{The Theory of Open Quantum Systems}}}}\ (\bibinfo  {publisher} {Oxford University Press},\ \bibinfo {address} {Oxford},\ \bibinfo {year} {2006})\BibitemShut {NoStop}%
\bibitem [{\citenamefont {Lindblad}(1976)}]{Lindblad1976b}%
  \BibitemOpen
  \bibfield  {author} {\bibinfo {author} {\bibfnamefont {G.}~\bibnamefont {Lindblad}},\ }\href {https://doi.org/10.1007/BF01608499} {\bibfield  {journal} {\bibinfo  {journal} {Commun. Math. Phys.}\ }\textbf {\bibinfo {volume} {48}},\ \bibinfo {pages} {119} (\bibinfo {year} {1976})}\BibitemShut {NoStop}%
\bibitem [{\citenamefont {Gorini}\ \emph {et~al.}(1976)\citenamefont {Gorini}, \citenamefont {Kossakowski},\ and\ \citenamefont {Sudarshan}}]{Gorini1976}%
  \BibitemOpen
  \bibfield  {author} {\bibinfo {author} {\bibfnamefont {V.}~\bibnamefont {Gorini}}, \bibinfo {author} {\bibfnamefont {A.}~\bibnamefont {Kossakowski}},\ and\ \bibinfo {author} {\bibfnamefont {E.~C.~G.}\ \bibnamefont {Sudarshan}},\ }\href {https://doi.org/10.1063/1.522979} {\bibfield  {journal} {\bibinfo  {journal} {J. Math. Phys.}\ }\textbf {\bibinfo {volume} {17}},\ \bibinfo {pages} {821} (\bibinfo {year} {1976})}\BibitemShut {NoStop}%
\bibitem [{\citenamefont {Pausch}\ \emph {et~al.}(2024)\citenamefont {Pausch}, \citenamefont {Damanet}, \citenamefont {Bastin},\ and\ \citenamefont {Martin}}]{Pausch2024}%
  \BibitemOpen
  \bibfield  {author} {\bibinfo {author} {\bibfnamefont {L.}~\bibnamefont {Pausch}}, \bibinfo {author} {\bibfnamefont {F.}~\bibnamefont {Damanet}}, \bibinfo {author} {\bibfnamefont {T.}~\bibnamefont {Bastin}},\ and\ \bibinfo {author} {\bibfnamefont {J.}~\bibnamefont {Martin}},\ }\href {https://doi.org/10.1103/PhysRevA.110.062208} {\bibfield  {journal} {\bibinfo  {journal} {Phys. Rev. A}\ }\textbf {\bibinfo {volume} {110}},\ \bibinfo {pages} {062208} (\bibinfo {year} {2024})}\BibitemShut {NoStop}%
\bibitem [{\citenamefont {Ferreira}\ and\ \citenamefont {Ribeiro}(2019)}]{FerreiraL2019}%
  \BibitemOpen
  \bibfield  {author} {\bibinfo {author} {\bibfnamefont {J.~a.~S.}\ \bibnamefont {Ferreira}}\ and\ \bibinfo {author} {\bibfnamefont {P.}~\bibnamefont {Ribeiro}},\ }\href {https://doi.org/10.1103/PhysRevB.100.184422} {\bibfield  {journal} {\bibinfo  {journal} {Phys. Rev. B}\ }\textbf {\bibinfo {volume} {100}},\ \bibinfo {pages} {184422} (\bibinfo {year} {2019})}\BibitemShut {NoStop}%
\bibitem [{\citenamefont {Ma}\ \emph {et~al.}(2020)\citenamefont {Ma}, \citenamefont {Ding},\ and\ \citenamefont {Yu}}]{Ma2020P}%
  \BibitemOpen
  \bibfield  {author} {\bibinfo {author} {\bibfnamefont {Y.-H.}\ \bibnamefont {Ma}}, \bibinfo {author} {\bibfnamefont {Q.-Z.}\ \bibnamefont {Ding}},\ and\ \bibinfo {author} {\bibfnamefont {T.}~\bibnamefont {Yu}},\ }\href {https://doi.org/10.1103/PhysRevA.101.022327} {\bibfield  {journal} {\bibinfo  {journal} {Phys. Rev. A}\ }\textbf {\bibinfo {volume} {101}},\ \bibinfo {pages} {022327} (\bibinfo {year} {2020})}\BibitemShut {NoStop}%
\bibitem [{\citenamefont {Lee}\ \emph {et~al.}(2014)\citenamefont {Lee}, \citenamefont {Chan},\ and\ \citenamefont {Yelin}}]{Lee2014D}%
  \BibitemOpen
  \bibfield  {author} {\bibinfo {author} {\bibfnamefont {T.~E.}\ \bibnamefont {Lee}}, \bibinfo {author} {\bibfnamefont {C.-K.}\ \bibnamefont {Chan}},\ and\ \bibinfo {author} {\bibfnamefont {S.~F.}\ \bibnamefont {Yelin}},\ }\href {https://doi.org/10.1103/PhysRevA.90.052109} {\bibfield  {journal} {\bibinfo  {journal} {Phys. Rev. A}\ }\textbf {\bibinfo {volume} {90}},\ \bibinfo {pages} {052109} (\bibinfo {year} {2014})}\BibitemShut {NoStop}%
\bibitem [{\citenamefont {Morrison}\ and\ \citenamefont {Parkins}(2008{\natexlab{a}})}]{Morrison2008D}%
  \BibitemOpen
  \bibfield  {author} {\bibinfo {author} {\bibfnamefont {S.}~\bibnamefont {Morrison}}\ and\ \bibinfo {author} {\bibfnamefont {A.~S.}\ \bibnamefont {Parkins}},\ }\href {https://doi.org/10.1103/PhysRevLett.100.040403} {\bibfield  {journal} {\bibinfo  {journal} {Phys. Rev. Lett.}\ }\textbf {\bibinfo {volume} {100}},\ \bibinfo {pages} {040403} (\bibinfo {year} {2008}{\natexlab{a}})}\BibitemShut {NoStop}%
\bibitem [{Note1()}]{Note1}%
  \BibitemOpen
  \bibinfo {note} {Given the quadratic form of the Hamiltonian \protect \eqref {H_S}, the ground state in the thermodynamic limit $N \to \infty $ is Gaussian and therefore a coherent or squeezed state, since it is pure. It is coherent only when $V/h = 0$, as a spin-coherent state can only be an eigenstate of \protect \eqref {H_S} when $H_{\protect \mathrm {S}} = hJ_z$. For $V/h \neq 0$, the state is therefore squeezed.}\BibitemShut {Stop}%
\bibitem [{\citenamefont {Nigro}(2019)}]{Nigro2019Uniqueness}%
  \BibitemOpen
  \bibfield  {author} {\bibinfo {author} {\bibfnamefont {D.}~\bibnamefont {Nigro}},\ }\href {https://doi.org/10.1088/1742-5468/ab0c1c} {\bibfield  {journal} {\bibinfo  {journal} {J. Stat. Mech.}\ }\textbf {\bibinfo {volume} {2019}},\ \bibinfo {pages} {043202} (\bibinfo {year} {2019})}\BibitemShut {NoStop}%
\bibitem [{\citenamefont {Morrison}\ and\ \citenamefont {Parkins}(2008{\natexlab{b}})}]{Morrison2008C}%
  \BibitemOpen
  \bibfield  {author} {\bibinfo {author} {\bibfnamefont {S.}~\bibnamefont {Morrison}}\ and\ \bibinfo {author} {\bibfnamefont {A.~S.}\ \bibnamefont {Parkins}},\ }\href {https://doi.org/10.1103/PhysRevA.77.043810} {\bibfield  {journal} {\bibinfo  {journal} {Phys. Rev. A}\ }\textbf {\bibinfo {volume} {77}},\ \bibinfo {pages} {043810} (\bibinfo {year} {2008}{\natexlab{b}})}\BibitemShut {NoStop}%
\bibitem [{\citenamefont {Farokh~Mivehvar}\ and\ \citenamefont {Ritsch}(2021)}]{Mivehvar2021Cavity}%
  \BibitemOpen
  \bibfield  {author} {\bibinfo {author} {\bibfnamefont {T.~D.}\ \bibnamefont {Farokh~Mivehvar}, \bibfnamefont {Francesco~Piazza}}\ and\ \bibinfo {author} {\bibfnamefont {H.}~\bibnamefont {Ritsch}},\ }\href {https://doi.org/10.1080/00018732.2021.1969727} {\bibfield  {journal} {\bibinfo  {journal} {Advances in Physics}\ }\textbf {\bibinfo {volume} {70}},\ \bibinfo {pages} {1} (\bibinfo {year} {2021})}\BibitemShut {NoStop}%
\bibitem [{\citenamefont {Blais}\ \emph {et~al.}(2021)\citenamefont {Blais}, \citenamefont {Grimsmo}, \citenamefont {Girvin},\ and\ \citenamefont {Wallraff}}]{Blais2021Circuit}%
  \BibitemOpen
  \bibfield  {author} {\bibinfo {author} {\bibfnamefont {A.}~\bibnamefont {Blais}}, \bibinfo {author} {\bibfnamefont {A.~L.}\ \bibnamefont {Grimsmo}}, \bibinfo {author} {\bibfnamefont {S.~M.}\ \bibnamefont {Girvin}},\ and\ \bibinfo {author} {\bibfnamefont {A.}~\bibnamefont {Wallraff}},\ }\href {https://doi.org/10.1103/RevModPhys.93.025005} {\bibfield  {journal} {\bibinfo  {journal} {Rev. Mod. Phys.}\ }\textbf {\bibinfo {volume} {93}},\ \bibinfo {pages} {025005} (\bibinfo {year} {2021})}\BibitemShut {NoStop}%
\bibitem [{\citenamefont {Imamoglu}(1994)}]{Imamoglu1994Stochastic}%
  \BibitemOpen
  \bibfield  {author} {\bibinfo {author} {\bibfnamefont {A.}~\bibnamefont {Imamoglu}},\ }\href {https://doi.org/10.1103/PhysRevA.50.3650} {\bibfield  {journal} {\bibinfo  {journal} {Phys. Rev. A}\ }\textbf {\bibinfo {volume} {50}},\ \bibinfo {pages} {3650} (\bibinfo {year} {1994})}\BibitemShut {NoStop}%
\bibitem [{\citenamefont {Dalton}\ \emph {et~al.}(2001)\citenamefont {Dalton}, \citenamefont {Barnett},\ and\ \citenamefont {Garraway}}]{Dalton2001Theory}%
  \BibitemOpen
  \bibfield  {author} {\bibinfo {author} {\bibfnamefont {B.~J.}\ \bibnamefont {Dalton}}, \bibinfo {author} {\bibfnamefont {S.~M.}\ \bibnamefont {Barnett}},\ and\ \bibinfo {author} {\bibfnamefont {B.~M.}\ \bibnamefont {Garraway}},\ }\href {https://doi.org/10.1103/PhysRevA.64.053813} {\bibfield  {journal} {\bibinfo  {journal} {Phys. Rev. A}\ }\textbf {\bibinfo {volume} {64}},\ \bibinfo {pages} {053813} (\bibinfo {year} {2001})}\BibitemShut {NoStop}%
\bibitem [{\citenamefont {Garraway}(1997)}]{Garraway1997Nonperturbative}%
  \BibitemOpen
  \bibfield  {author} {\bibinfo {author} {\bibfnamefont {B.~M.}\ \bibnamefont {Garraway}},\ }\href {https://doi.org/10.1103/PhysRevA.55.2290} {\bibfield  {journal} {\bibinfo  {journal} {Phys. Rev. A}\ }\textbf {\bibinfo {volume} {55}},\ \bibinfo {pages} {2290} (\bibinfo {year} {1997})}\BibitemShut {NoStop}%
\bibitem [{\citenamefont {Pleasance}\ \emph {et~al.}(2020)\citenamefont {Pleasance}, \citenamefont {Garraway},\ and\ \citenamefont {Petruccione}}]{Pleasance2020Generalized}%
  \BibitemOpen
  \bibfield  {author} {\bibinfo {author} {\bibfnamefont {G.}~\bibnamefont {Pleasance}}, \bibinfo {author} {\bibfnamefont {B.~M.}\ \bibnamefont {Garraway}},\ and\ \bibinfo {author} {\bibfnamefont {F.}~\bibnamefont {Petruccione}},\ }\href {https://doi.org/10.1103/PhysRevResearch.2.043058} {\bibfield  {journal} {\bibinfo  {journal} {Phys. Rev. Research}\ }\textbf {\bibinfo {volume} {2}},\ \bibinfo {pages} {043058} (\bibinfo {year} {2020})}\BibitemShut {NoStop}%
\bibitem [{\citenamefont {Mazzola}\ \emph {et~al.}(2009)\citenamefont {Mazzola}, \citenamefont {Maniscalco}, \citenamefont {Piilo}, \citenamefont {Suominen},\ and\ \citenamefont {Garraway}}]{Mazzola2009Pseudomodes}%
  \BibitemOpen
  \bibfield  {author} {\bibinfo {author} {\bibfnamefont {L.}~\bibnamefont {Mazzola}}, \bibinfo {author} {\bibfnamefont {S.}~\bibnamefont {Maniscalco}}, \bibinfo {author} {\bibfnamefont {J.}~\bibnamefont {Piilo}}, \bibinfo {author} {\bibfnamefont {K.-A.}\ \bibnamefont {Suominen}},\ and\ \bibinfo {author} {\bibfnamefont {B.~M.}\ \bibnamefont {Garraway}},\ }\href {https://doi.org/10.1103/PhysRevA.80.012104} {\bibfield  {journal} {\bibinfo  {journal} {Phys. Rev. A}\ }\textbf {\bibinfo {volume} {80}},\ \bibinfo {pages} {012104} (\bibinfo {year} {2009})}\BibitemShut {NoStop}%
\bibitem [{\citenamefont {Yang}\ \emph {et~al.}(2012)\citenamefont {Yang}, \citenamefont {Miao},\ and\ \citenamefont {Chen}}]{Yang2012Nonadiabatic}%
  \BibitemOpen
  \bibfield  {author} {\bibinfo {author} {\bibfnamefont {H.}~\bibnamefont {Yang}}, \bibinfo {author} {\bibfnamefont {H.}~\bibnamefont {Miao}},\ and\ \bibinfo {author} {\bibfnamefont {Y.}~\bibnamefont {Chen}},\ }\href {https://doi.org/10.1103/PhysRevA.85.040101} {\bibfield  {journal} {\bibinfo  {journal} {Phys. Rev. A}\ }\textbf {\bibinfo {volume} {85}},\ \bibinfo {pages} {040101} (\bibinfo {year} {2012})}\BibitemShut {NoStop}%
\bibitem [{\citenamefont {Breuer}(2004)}]{Breuer2004Genuine}%
  \BibitemOpen
  \bibfield  {author} {\bibinfo {author} {\bibfnamefont {H.-P.}\ \bibnamefont {Breuer}},\ }\href {https://doi.org/10.1103/PhysRevA.70.012106} {\bibfield  {journal} {\bibinfo  {journal} {Phys. Rev. A}\ }\textbf {\bibinfo {volume} {70}},\ \bibinfo {pages} {012106} (\bibinfo {year} {2004})}\BibitemShut {NoStop}%
\bibitem [{\citenamefont {Barchielli}\ \emph {et~al.}(2010)\citenamefont {Barchielli}, \citenamefont {Pellegrini},\ and\ \citenamefont {Petruccione}}]{Barchielli2010Stochastic}%
  \BibitemOpen
  \bibfield  {author} {\bibinfo {author} {\bibfnamefont {A.}~\bibnamefont {Barchielli}}, \bibinfo {author} {\bibfnamefont {C.}~\bibnamefont {Pellegrini}},\ and\ \bibinfo {author} {\bibfnamefont {F.}~\bibnamefont {Petruccione}},\ }\href {https://doi.org/10.1209/0295-5075/91/24001} {\bibfield  {journal} {\bibinfo  {journal} {{EPL} (Europhysics Letters)}\ }\textbf {\bibinfo {volume} {91}},\ \bibinfo {pages} {24001} (\bibinfo {year} {2010})}\BibitemShut {NoStop}%
\bibitem [{\citenamefont {Tanimura}(2020)}]{Tanimura2020}%
  \BibitemOpen
  \bibfield  {author} {\bibinfo {author} {\bibfnamefont {Y.}~\bibnamefont {Tanimura}},\ }\href {https://doi.org/10.1063/5.0011599} {\bibfield  {journal} {\bibinfo  {journal} {The Journal of Chemical Physics}\ }\textbf {\bibinfo {volume} {153}},\ \bibinfo {pages} {020901} (\bibinfo {year} {2020})}\BibitemShut {NoStop}%
\bibitem [{\citenamefont {Tanimura}\ and\ \citenamefont {Kubo}(1989)}]{Tanimura89}%
  \BibitemOpen
  \bibfield  {author} {\bibinfo {author} {\bibfnamefont {Y.}~\bibnamefont {Tanimura}}\ and\ \bibinfo {author} {\bibfnamefont {R.}~\bibnamefont {Kubo}},\ }\href {https://doi.org/10.1143/JPSJ.58.1199} {\bibfield  {journal} {\bibinfo  {journal} {Journal of the Physical Society of Japan}\ }\textbf {\bibinfo {volume} {58}},\ \bibinfo {pages} {1199} (\bibinfo {year} {1989})}\BibitemShut {NoStop}%
\bibitem [{\citenamefont {Buča}\ and\ \citenamefont {Prosen}(2012)}]{Buca2012}%
  \BibitemOpen
  \bibfield  {author} {\bibinfo {author} {\bibfnamefont {B.}~\bibnamefont {Buča}}\ and\ \bibinfo {author} {\bibfnamefont {T.}~\bibnamefont {Prosen}},\ }\href {https://doi.org/10.1088/1367-2630/14/7/073007} {\bibfield  {journal} {\bibinfo  {journal} {New Journal of Physics}\ }\textbf {\bibinfo {volume} {14}},\ \bibinfo {pages} {073007} (\bibinfo {year} {2012})}\BibitemShut {NoStop}%
\bibitem [{\citenamefont {Albert}\ and\ \citenamefont {Jiang}(2014)}]{Albert2014S}%
  \BibitemOpen
  \bibfield  {author} {\bibinfo {author} {\bibfnamefont {V.~V.}\ \bibnamefont {Albert}}\ and\ \bibinfo {author} {\bibfnamefont {L.}~\bibnamefont {Jiang}},\ }\href {https://doi.org/10.1103/PhysRevA.89.022118} {\bibfield  {journal} {\bibinfo  {journal} {Phys. Rev. A}\ }\textbf {\bibinfo {volume} {89}},\ \bibinfo {pages} {022118} (\bibinfo {year} {2014})}\BibitemShut {NoStop}%
\bibitem [{\citenamefont {Minganti}\ \emph {et~al.}(2018)\citenamefont {Minganti}, \citenamefont {Biella}, \citenamefont {Bartolo},\ and\ \citenamefont {Ciuti}}]{Minganti2018Spectral}%
  \BibitemOpen
  \bibfield  {author} {\bibinfo {author} {\bibfnamefont {F.}~\bibnamefont {Minganti}}, \bibinfo {author} {\bibfnamefont {A.}~\bibnamefont {Biella}}, \bibinfo {author} {\bibfnamefont {N.}~\bibnamefont {Bartolo}},\ and\ \bibinfo {author} {\bibfnamefont {C.}~\bibnamefont {Ciuti}},\ }\href {https://doi.org/10.1103/PhysRevA.98.042118} {\bibfield  {journal} {\bibinfo  {journal} {Phys. Rev. A}\ }\textbf {\bibinfo {volume} {98}},\ \bibinfo {pages} {042118} (\bibinfo {year} {2018})}\BibitemShut {NoStop}%
\bibitem [{\citenamefont {Hepp}\ and\ \citenamefont {Lieb}(1973)}]{Hepp1973O}%
  \BibitemOpen
  \bibfield  {author} {\bibinfo {author} {\bibfnamefont {K.}~\bibnamefont {Hepp}}\ and\ \bibinfo {author} {\bibfnamefont {E.~H.}\ \bibnamefont {Lieb}},\ }\href {https://doi.org/https://doi.org/10.1016/0003-4916(73)90039-0} {\bibfield  {journal} {\bibinfo  {journal} {Annals of Physics}\ }\textbf {\bibinfo {volume} {76}},\ \bibinfo {pages} {360} (\bibinfo {year} {1973})}\BibitemShut {NoStop}%
\bibitem [{\citenamefont {Minganti}\ \emph {et~al.}(2021)\citenamefont {Minganti}, \citenamefont {Arkhipov}, \citenamefont {Miranowicz},\ and\ \citenamefont {Nori}}]{Minganti2021Continuous}%
  \BibitemOpen
  \bibfield  {author} {\bibinfo {author} {\bibfnamefont {F.}~\bibnamefont {Minganti}}, \bibinfo {author} {\bibfnamefont {I.~I.}\ \bibnamefont {Arkhipov}}, \bibinfo {author} {\bibfnamefont {A.}~\bibnamefont {Miranowicz}},\ and\ \bibinfo {author} {\bibfnamefont {F.}~\bibnamefont {Nori}},\ }\href {https://doi.org/10.1088/1367-2630/ac3db8} {\bibfield  {journal} {\bibinfo  {journal} {New Journal of Physics}\ }\textbf {\bibinfo {volume} {23}},\ \bibinfo {pages} {122001} (\bibinfo {year} {2021})}\BibitemShut {NoStop}%
\bibitem [{\citenamefont {Huber}\ \emph {et~al.}(2020)\citenamefont {Huber}, \citenamefont {Kirton},\ and\ \citenamefont {Rabl}}]{Huber2020}%
  \BibitemOpen
  \bibfield  {author} {\bibinfo {author} {\bibfnamefont {J.}~\bibnamefont {Huber}}, \bibinfo {author} {\bibfnamefont {P.}~\bibnamefont {Kirton}},\ and\ \bibinfo {author} {\bibfnamefont {P.}~\bibnamefont {Rabl}},\ }\href {https://doi.org/10.1103/PhysRevA.102.012219} {\bibfield  {journal} {\bibinfo  {journal} {Phys. Rev. A}\ }\textbf {\bibinfo {volume} {102}},\ \bibinfo {pages} {012219} (\bibinfo {year} {2020})}\BibitemShut {NoStop}%
\bibitem [{\citenamefont {Gu}\ \emph {et~al.}(2024)\citenamefont {Gu}, \citenamefont {Wang},\ and\ \citenamefont {Wang}}]{Gu2024S}%
  \BibitemOpen
  \bibfield  {author} {\bibinfo {author} {\bibfnamefont {D.}~\bibnamefont {Gu}}, \bibinfo {author} {\bibfnamefont {Z.}~\bibnamefont {Wang}},\ and\ \bibinfo {author} {\bibfnamefont {Z.}~\bibnamefont {Wang}},\ }\href {https://arxiv.org/abs/2406.19381} {\bibinfo {title} {Spontaneous symmetry breaking in open quantum systems: strong, weak, and strong-to-weak}} (\bibinfo {year} {2024}),\ \Eprint {https://arxiv.org/abs/2406.19381} {arXiv:2406.19381 [quant-ph]} \BibitemShut {NoStop}%
\bibitem [{\citenamefont {Baksic}\ and\ \citenamefont {Ciuti}(2014)}]{Baksic2014C}%
  \BibitemOpen
  \bibfield  {author} {\bibinfo {author} {\bibfnamefont {A.}~\bibnamefont {Baksic}}\ and\ \bibinfo {author} {\bibfnamefont {C.}~\bibnamefont {Ciuti}},\ }\href {https://doi.org/10.1103/PhysRevLett.112.173601} {\bibfield  {journal} {\bibinfo  {journal} {Phys. Rev. Lett.}\ }\textbf {\bibinfo {volume} {112}},\ \bibinfo {pages} {173601} (\bibinfo {year} {2014})}\BibitemShut {NoStop}%
\bibitem [{Note2()}]{Note2}%
  \BibitemOpen
  \bibinfo {note} {Recall that the finite-size results displayed in Fig.~\ref {fig:DSSB} are obtained via the HEOM Liouvillian~(\ref {HEOM}), rather than from the Markovian embedding [Eq.~\protect \eqref {master_LMG_tot}] which includes the pseudomode in the system. As a consequence, the solution includes auxiliary operators not directly relevant here, and one must select the $(0, 0)$ component of the appropriate eigenvector, as covered in detail in Ref.~\cite {Debecker2024S}}\BibitemShut {NoStop}%
\bibitem [{\citenamefont {Kitagawa}\ and\ \citenamefont {Ueda}(1993)}]{Kitagawa1993S}%
  \BibitemOpen
  \bibfield  {author} {\bibinfo {author} {\bibfnamefont {M.}~\bibnamefont {Kitagawa}}\ and\ \bibinfo {author} {\bibfnamefont {M.}~\bibnamefont {Ueda}},\ }\href {https://doi.org/10.1103/PhysRevA.47.5138} {\bibfield  {journal} {\bibinfo  {journal} {Phys. Rev. A}\ }\textbf {\bibinfo {volume} {47}},\ \bibinfo {pages} {5138} (\bibinfo {year} {1993})}\BibitemShut {NoStop}%
\bibitem [{\citenamefont {Korbicz}\ \emph {et~al.}(2005)\citenamefont {Korbicz}, \citenamefont {Cirac},\ and\ \citenamefont {Lewenstein}}]{Korbicz2005S}%
  \BibitemOpen
  \bibfield  {author} {\bibinfo {author} {\bibfnamefont {J.~K.}\ \bibnamefont {Korbicz}}, \bibinfo {author} {\bibfnamefont {J.~I.}\ \bibnamefont {Cirac}},\ and\ \bibinfo {author} {\bibfnamefont {M.}~\bibnamefont {Lewenstein}},\ }\href {https://doi.org/10.1103/PhysRevLett.95.120502} {\bibfield  {journal} {\bibinfo  {journal} {Phys. Rev. Lett.}\ }\textbf {\bibinfo {volume} {95}},\ \bibinfo {pages} {120502} (\bibinfo {year} {2005})}\BibitemShut {NoStop}%
\bibitem [{\citenamefont {Pezz\`e}\ \emph {et~al.}(2018)\citenamefont {Pezz\`e}, \citenamefont {Smerzi}, \citenamefont {Oberthaler}, \citenamefont {Schmied},\ and\ \citenamefont {Treutlein}}]{Pezze2018Q}%
  \BibitemOpen
  \bibfield  {author} {\bibinfo {author} {\bibfnamefont {L.}~\bibnamefont {Pezz\`e}}, \bibinfo {author} {\bibfnamefont {A.}~\bibnamefont {Smerzi}}, \bibinfo {author} {\bibfnamefont {M.~K.}\ \bibnamefont {Oberthaler}}, \bibinfo {author} {\bibfnamefont {R.}~\bibnamefont {Schmied}},\ and\ \bibinfo {author} {\bibfnamefont {P.}~\bibnamefont {Treutlein}},\ }\href {https://doi.org/10.1103/RevModPhys.90.035005} {\bibfield  {journal} {\bibinfo  {journal} {Rev. Mod. Phys.}\ }\textbf {\bibinfo {volume} {90}},\ \bibinfo {pages} {035005} (\bibinfo {year} {2018})}\BibitemShut {NoStop}%
\bibitem [{\citenamefont {Ma}\ \emph {et~al.}(2011)\citenamefont {Ma}, \citenamefont {Wang}, \citenamefont {Sun},\ and\ \citenamefont {Nori}}]{Ma2011Q}%
  \BibitemOpen
  \bibfield  {author} {\bibinfo {author} {\bibfnamefont {J.}~\bibnamefont {Ma}}, \bibinfo {author} {\bibfnamefont {X.}~\bibnamefont {Wang}}, \bibinfo {author} {\bibfnamefont {C.}~\bibnamefont {Sun}},\ and\ \bibinfo {author} {\bibfnamefont {F.}~\bibnamefont {Nori}},\ }\href {https://doi.org/https://doi.org/10.1016/j.physrep.2011.08.003} {\bibfield  {journal} {\bibinfo  {journal} {Physics Reports}\ }\textbf {\bibinfo {volume} {509}},\ \bibinfo {pages} {89} (\bibinfo {year} {2011})}\BibitemShut {NoStop}%
\bibitem [{\citenamefont {Prosen}(2008)}]{Prosen2008T}%
  \BibitemOpen
  \bibfield  {author} {\bibinfo {author} {\bibfnamefont {T.}~\bibnamefont {Prosen}},\ }\href {https://doi.org/10.1088/1367-2630/10/4/043026} {\bibfield  {journal} {\bibinfo  {journal} {New Journal of Physics}\ }\textbf {\bibinfo {volume} {10}},\ \bibinfo {pages} {043026} (\bibinfo {year} {2008})}\BibitemShut {NoStop}%
\bibitem [{\citenamefont {Prosen}\ and\ \citenamefont {Seligman}(2010)}]{Prosen2010Q}%
  \BibitemOpen
  \bibfield  {author} {\bibinfo {author} {\bibfnamefont {T.}~\bibnamefont {Prosen}}\ and\ \bibinfo {author} {\bibfnamefont {T.~H.}\ \bibnamefont {Seligman}},\ }\href {https://doi.org/10.1088/1751-8113/43/39/392004} {\bibfield  {journal} {\bibinfo  {journal} {Journal of Physics A: Mathematical and Theoretical}\ }\textbf {\bibinfo {volume} {43}},\ \bibinfo {pages} {392004} (\bibinfo {year} {2010})}\BibitemShut {NoStop}%
\bibitem [{\citenamefont {Ma}\ and\ \citenamefont {Wang}(2009)}]{Ma2009F}%
  \BibitemOpen
  \bibfield  {author} {\bibinfo {author} {\bibfnamefont {J.}~\bibnamefont {Ma}}\ and\ \bibinfo {author} {\bibfnamefont {X.}~\bibnamefont {Wang}},\ }\href {https://doi.org/10.1103/PhysRevA.80.012318} {\bibfield  {journal} {\bibinfo  {journal} {Phys. Rev. A}\ }\textbf {\bibinfo {volume} {80}},\ \bibinfo {pages} {012318} (\bibinfo {year} {2009})}\BibitemShut {NoStop}%
\bibitem [{\citenamefont {Dusuel}\ and\ \citenamefont {Vidal}(2005)}]{Dusuel2005C}%
  \BibitemOpen
  \bibfield  {author} {\bibinfo {author} {\bibfnamefont {S.}~\bibnamefont {Dusuel}}\ and\ \bibinfo {author} {\bibfnamefont {J.}~\bibnamefont {Vidal}},\ }\href {https://doi.org/10.1103/PhysRevB.71.224420} {\bibfield  {journal} {\bibinfo  {journal} {Phys. Rev. B}\ }\textbf {\bibinfo {volume} {71}},\ \bibinfo {pages} {224420} (\bibinfo {year} {2005})}\BibitemShut {NoStop}%
\bibitem [{\citenamefont {Emary}\ and\ \citenamefont {Brandes}(2003)}]{Emary2003C}%
  \BibitemOpen
  \bibfield  {author} {\bibinfo {author} {\bibfnamefont {C.}~\bibnamefont {Emary}}\ and\ \bibinfo {author} {\bibfnamefont {T.}~\bibnamefont {Brandes}},\ }\href {https://doi.org/10.1103/PhysRevE.67.066203} {\bibfield  {journal} {\bibinfo  {journal} {Phys. Rev. E}\ }\textbf {\bibinfo {volume} {67}},\ \bibinfo {pages} {066203} (\bibinfo {year} {2003})}\BibitemShut {NoStop}%
\bibitem [{\citenamefont {Bu\ifmmode~\check{c}\else \v{c}\fi{}a}\ and\ \citenamefont {Jaksch}(2019)}]{Buca2019D}%
  \BibitemOpen
  \bibfield  {author} {\bibinfo {author} {\bibfnamefont {B.}~\bibnamefont {Bu\ifmmode~\check{c}\else \v{c}\fi{}a}}\ and\ \bibinfo {author} {\bibfnamefont {D.}~\bibnamefont {Jaksch}},\ }\href {https://doi.org/10.1103/PhysRevLett.123.260401} {\bibfield  {journal} {\bibinfo  {journal} {Phys. Rev. Lett.}\ }\textbf {\bibinfo {volume} {123}},\ \bibinfo {pages} {260401} (\bibinfo {year} {2019})}\BibitemShut {NoStop}%
\bibitem [{\citenamefont {Fan}\ and\ \citenamefont {Jia}(2023)}]{Fan2023C}%
  \BibitemOpen
  \bibfield  {author} {\bibinfo {author} {\bibfnamefont {J.}~\bibnamefont {Fan}}\ and\ \bibinfo {author} {\bibfnamefont {S.}~\bibnamefont {Jia}},\ }\href {https://doi.org/10.1103/PhysRevA.107.033711} {\bibfield  {journal} {\bibinfo  {journal} {Phys. Rev. A}\ }\textbf {\bibinfo {volume} {107}},\ \bibinfo {pages} {033711} (\bibinfo {year} {2023})}\BibitemShut {NoStop}%
\bibitem [{\citenamefont {Dimer}\ \emph {et~al.}(2007)\citenamefont {Dimer}, \citenamefont {Estienne}, \citenamefont {Parkins},\ and\ \citenamefont {Carmichael}}]{Dimer2007Jan}%
  \BibitemOpen
  \bibfield  {author} {\bibinfo {author} {\bibfnamefont {F.}~\bibnamefont {Dimer}}, \bibinfo {author} {\bibfnamefont {B.}~\bibnamefont {Estienne}}, \bibinfo {author} {\bibfnamefont {A.~S.}\ \bibnamefont {Parkins}},\ and\ \bibinfo {author} {\bibfnamefont {H.~J.}\ \bibnamefont {Carmichael}},\ }\href {https://doi.org/10.1103/PhysRevA.75.013804} {\bibfield  {journal} {\bibinfo  {journal} {Phys. Rev. A}\ }\textbf {\bibinfo {volume} {75}},\ \bibinfo {pages} {013804} (\bibinfo {year} {2007})}\BibitemShut {NoStop}%
\bibitem [{\citenamefont {Morrison}\ and\ \citenamefont {Parkins}(2008{\natexlab{c}})}]{Morrison2008}%
  \BibitemOpen
  \bibfield  {author} {\bibinfo {author} {\bibfnamefont {S.}~\bibnamefont {Morrison}}\ and\ \bibinfo {author} {\bibfnamefont {A.~S.}\ \bibnamefont {Parkins}},\ }\href {https://doi.org/10.1103/physrevlett.100.040403} {\bibfield  {journal} {\bibinfo  {journal} {Phys. Rev. Lett.}\ }\textbf {\bibinfo {volume} {100}},\ \bibinfo {pages} {040403} (\bibinfo {year} {2008}{\natexlab{c}})}\BibitemShut {NoStop}%
\bibitem [{Note3()}]{Note3}%
  \BibitemOpen
  \bibinfo {note} {Note that these parameters differ slightly from those presented in Ref.~\cite {Morrison2008C}, since there was a sign error in that publication and the authors of Ref.~\cite {Morrison2008C} omitted a contribution to $\omega _0$ that we are including.}\BibitemShut {Stop}%
\bibitem [{\citenamefont {Metcalf}\ and\ \citenamefont {van~der Straten}(1999)}]{MetcalfBook}%
  \BibitemOpen
  \bibfield  {author} {\bibinfo {author} {\bibfnamefont {H.~J.}\ \bibnamefont {Metcalf}}\ and\ \bibinfo {author} {\bibfnamefont {P.}~\bibnamefont {van~der Straten}},\ }\href {https://doi.org/10.1007/978-1-4612-1470-0} {\emph {\bibinfo {title} {Laser Cooling and Trapping}}}\ (\bibinfo  {publisher} {Springer New York},\ \bibinfo {year} {1999})\BibitemShut {NoStop}%
\bibitem [{\citenamefont {Azouit}\ \emph {et~al.}(2017)\citenamefont {Azouit}, \citenamefont {Chittaro}, \citenamefont {Sarlette},\ and\ \citenamefont {Rouchon}}]{Azouit2017}%
  \BibitemOpen
  \bibfield  {author} {\bibinfo {author} {\bibfnamefont {R.}~\bibnamefont {Azouit}}, \bibinfo {author} {\bibfnamefont {F.}~\bibnamefont {Chittaro}}, \bibinfo {author} {\bibfnamefont {A.}~\bibnamefont {Sarlette}},\ and\ \bibinfo {author} {\bibfnamefont {P.}~\bibnamefont {Rouchon}},\ }\href {https://doi.org/10.1088/2058-9565/aa7f3f} {\bibfield  {journal} {\bibinfo  {journal} {Quantum Science and Technology}\ }\textbf {\bibinfo {volume} {2}},\ \bibinfo {pages} {044011} (\bibinfo {year} {2017})}\BibitemShut {NoStop}%
\bibitem [{\citenamefont {Lieu}\ \emph {et~al.}(2020)\citenamefont {Lieu}, \citenamefont {Belyansky}, \citenamefont {Young}, \citenamefont {Lundgren}, \citenamefont {Albert},\ and\ \citenamefont {Gorshkov}}]{Lieu2020S}%
  \BibitemOpen
  \bibfield  {author} {\bibinfo {author} {\bibfnamefont {S.}~\bibnamefont {Lieu}}, \bibinfo {author} {\bibfnamefont {R.}~\bibnamefont {Belyansky}}, \bibinfo {author} {\bibfnamefont {J.~T.}\ \bibnamefont {Young}}, \bibinfo {author} {\bibfnamefont {R.}~\bibnamefont {Lundgren}}, \bibinfo {author} {\bibfnamefont {V.~V.}\ \bibnamefont {Albert}},\ and\ \bibinfo {author} {\bibfnamefont {A.~V.}\ \bibnamefont {Gorshkov}},\ }\href {https://doi.org/10.1103/PhysRevLett.125.240405} {\bibfield  {journal} {\bibinfo  {journal} {Phys. Rev. Lett.}\ }\textbf {\bibinfo {volume} {125}},\ \bibinfo {pages} {240405} (\bibinfo {year} {2020})}\BibitemShut {NoStop}%
\bibitem [{\citenamefont {Sala}\ \emph {et~al.}(2024)\citenamefont {Sala}, \citenamefont {Gopalakrishnan}, \citenamefont {Oshikawa},\ and\ \citenamefont {You}}]{Pablo2024S}%
  \BibitemOpen
  \bibfield  {author} {\bibinfo {author} {\bibfnamefont {P.}~\bibnamefont {Sala}}, \bibinfo {author} {\bibfnamefont {S.}~\bibnamefont {Gopalakrishnan}}, \bibinfo {author} {\bibfnamefont {M.}~\bibnamefont {Oshikawa}},\ and\ \bibinfo {author} {\bibfnamefont {Y.}~\bibnamefont {You}},\ }\href {https://doi.org/10.1103/PhysRevB.110.155150} {\bibfield  {journal} {\bibinfo  {journal} {Phys. Rev. B}\ }\textbf {\bibinfo {volume} {110}},\ \bibinfo {pages} {155150} (\bibinfo {year} {2024})}\BibitemShut {NoStop}%
\bibitem [{\citenamefont {Lessa}\ \emph {et~al.}(2025)\citenamefont {Lessa}, \citenamefont {Ma}, \citenamefont {Zhang}, \citenamefont {Bi}, \citenamefont {Cheng},\ and\ \citenamefont {Wang}}]{Lessa2025S}%
  \BibitemOpen
  \bibfield  {author} {\bibinfo {author} {\bibfnamefont {L.~A.}\ \bibnamefont {Lessa}}, \bibinfo {author} {\bibfnamefont {R.}~\bibnamefont {Ma}}, \bibinfo {author} {\bibfnamefont {J.-H.}\ \bibnamefont {Zhang}}, \bibinfo {author} {\bibfnamefont {Z.}~\bibnamefont {Bi}}, \bibinfo {author} {\bibfnamefont {M.}~\bibnamefont {Cheng}},\ and\ \bibinfo {author} {\bibfnamefont {C.}~\bibnamefont {Wang}},\ }\href {https://doi.org/10.1103/PRXQuantum.6.010344} {\bibfield  {journal} {\bibinfo  {journal} {PRX Quantum}\ }\textbf {\bibinfo {volume} {6}},\ \bibinfo {pages} {010344} (\bibinfo {year} {2025})}\BibitemShut {NoStop}%
\end{thebibliography}%

\end{document}